%% file: cas-dc-template.tex
\documentclass[a4paper,fleqn]{cas-dc}

\usepackage[numbers]{natbib}
\usepackage{amsmath}
\usepackage{bm}
\usepackage{colortbl}
\usepackage{xcolor}
\usepackage{color}
\usepackage{soul}
\usepackage{bbding}
\usepackage{multirow}
\usepackage{multicol}
\usepackage{booktabs}
\definecolor{light-gray}{gray}{0.90}
\definecolor{mydeepgreen}{RGB}{0,180,0}
\hypersetup{colorlinks=true, linkcolor=blue, anchorcolor=blue, citecolor=blue}

\usepackage{algorithm,algpseudocode,listings}
\definecolor{codegreen}{rgb}{0,0.6,0}
\definecolor{codegray}{rgb}{0.5,0.5,0.5}
\definecolor{codepurple}{rgb}{0.58,0,0.82}
\definecolor{backcolour}{rgb}{1.0,1.0,1.0}
\lstdefinestyle{mystyle}{
    backgroundcolor=\color{backcolour},
    commentstyle=\color{codegreen},
    keywordstyle=\color{magenta},
    numberstyle=\tiny\color{codegray},
    stringstyle=\color{codepurple},
    basicstyle=\ttfamily\scriptsize,
    breakatwhitespace=false,
    breaklines=true,
    captionpos=b,
    keepspaces=true,
    showspaces=false,
    showstringspaces=false,
    showtabs=false,
    tabsize=2
}
\def\tsc#1{\csdef{#1}{\textsc{\lowercase{#1}}\xspace}}
\tsc{WGM}
\tsc{QE}
\tsc{EP}
\tsc{PMS}
\tsc{BEC}
\tsc{DE}

\begin{document}
\let\printorcid\relax
\let\WriteBookmarks\relax
\def\floatpagepagefraction{1}
\def\textpagefraction{.001}

\shorttitle{SPIRONet: Spatial-Frequency Learning and Graph-based Channel Interaction Network for Vessel Segmentation}

\shortauthors{D.-X. Huang et~al.}

\title [mode = title]{SPIRONet: Spatial-Frequency Learning and Graph-based Channel Interaction Network for Vessel Segmentation}

\author[1,2]{De-Xing Huang}
\ead{huangdexing2022@ia.ac.cn}
\credit{Data curation, Methodology, Writing - original draft, Writing - review \& editing}

\author[1,2]{Xiao-Hu Zhou}\cormark[1]
\ead{xiaohu.zhou@ia.ac.cn}
\credit{Conceptualization, Funding acquisition, Writing - review \& editing}

\author[1]{Xiao-Liang Xie}
\credit{Funding acquisition}

\author[1]{Shi-Qi Liu}
\credit{Resources}

\author[1,2]{Shuangyi Wang}
\credit{Resources}

\author[1]{Zhen-Qiu Feng}
\credit{Resources}

\author[1]{Mei-Jiang Gui}
\credit{Resources, Funding acquisition}

\author[1,2]{Hao Li}
\credit{Methodology, Writing - review \& editing}

\author[1,2]{Tian-Yu Xiang}
\credit{Methodology, Writing - review \& editing}

\author[1,2]{Bo-Xian Yao}
\credit{Writing - review \& editing}

\author[1,2]{Zeng-Guang Hou}\cormark[1]
\ead{zengguang.hou@ia.ac.cn}
\credit{Supervision, Funding acquisition, Writing - review \& editing}

\affiliation[1]{organization={State Key Laboratory of Multimodal Artificial Intelligence Systems, Institute of Automation, Chinese Academy of Sciences},
city={Beijing},
postcode={100190},
country={China}}

\affiliation[2]{organization={School of Artificial Intelligence, University of Chinese Academy of Sciences},
city={Beijing},
postcode={100049},
country={China}}

\nonumnote{$^*$Corresponding authors.}

\begin{abstract}
Automatic vessel segmentation plays a pivotal role in the development of next-generation interventional navigation systems for surgical robotics. However, current approaches still suffer from suboptimal segmentation performance under challenging intraoperative conditions, such as low-signal-to-noise ratio (SNR), small or slender vessels, and strong interference. In this study, a novel \underline{\textbf{SP}}atial-frequency learning and graph-based channel \underline{\textbf{I}}nte\underline{\textbf{R}}acti\underline{\textbf{O}}n \underline{\textbf{Net}}work (\textbf{SPIRONet}) is proposed to address the above issues. To address low-SNR vessel appearance and small or slender branches, dual spatial-frequency encoders are utilized, where the frequency encoder captures global vessel continuity that is less affected by local noise fluctuations, while the spatial encoder preserves fine vessel details. A cross-attention fusion module is further introduced to adaptively integrate this complementary spatial and frequency information. Moreover, to suppress interference from non-target vessels and vessel-like structures, a graph-based channel interaction module is designed to model channel-wise correlations, enhancing consistent vessel-related responses while suppressing task-irrelevant activations. Extensive experimental results on five challenging datasets demonstrate that the proposed method achieves competitive and consistently strong performance compared with existing methods. For example, SPIRONet achieves IoU improvements of $+0.87\%$, $+0.52\%$, $+0.23\%$, $+1.39\%$, and $+2.22\%$ over the strongest competing methods on \textit{CADSA}, \textit{CAXF}, \textit{DCA1}, \textit{XCAD}, and \textit{ARCADE}, respectively. Moreover, SPIRONet achieves an inference speed of 21 FPS with a 512$\times$512 input size, meeting the real-time requirements of interventional scenarios ($6-12$ FPS). These promising results indicate SPIRONet's potential for integration into interventional navigation systems. Code is available at \href{https://github.com/Dxhuang-CASIA/SPIRONet}{\tt \color{blue}https://github.com/Dxhuang-CASIA/SPIRONet}.
\end{abstract}



\maketitle

\section{Introduction} \label{sec:s1}
\input{sec1}

\section{Related Work} \label{sec:s2}
\input{sec2}

\section{Method} \label{sec:s3}
\input{sec3}

\section{Experimental Setup} \label{sec:s4}
\input{sec4}

\section{Results} \label{sec:s5}
\input{sec5}

\section{In-Depth Analysis} \label{sec:s6}
\input{sec6}

\section{Conclusion} \label{sec:s7}
\input{sec7}

\printcredits

\section*{Acknowledgments}
This work was supported in part by the National Key Research and Development Program of China under Grant 2023YFC2415100, in part by the National Natural Science Foundation of China under Grant 62373351, Grant 62503474, Grant 82327801, Grant 62303463, in part by the Chinese Academy of Sciences Project for Research Instrument Developmentunder under Grant No. PTYQ2026YZ0072, Young Scientists in Basic Research under Grant No. YSBR-104, in part by the Beijing Natural Science Foundation under Grant F252068, Grant 4254107, in part by Beijing Nova Program under Grant 20250484813, in part by China Postdoctoral Science Foundation under Grant 2024M763535, in part by the Postdoctoral Fellowship Program of CPSF under Grant GZC20251170, and in part by Xiaomi Young Talents Program/Xiaomi Foundation.

\section*{Declaration of Competing Interest}
The authors declare that they have no known competing financial interests or personal relationships that could have appeared to influence the work reported in this paper.

\section*{Data Availability}
Data will be made available upon reasonable request.

\bibliographystyle{model1-num-names}

\bibliography{cas-refs}

\end{document}

%% file: sec1.tex
Cardiovascular diseases are major contributors to global morbidity and mortality rates~\cite{sacco2016heart},~\cite{roth2020global}. Intravascular interventions have gained considerable attention because they are minimally invasive and enable rapid postoperative recovery~\cite{wan2022symptomatic},~\cite{langer2016minimally}. These procedures require physicians to deliver instruments (\textit{e.g.,} guidewires, catheters, or balloons) precisely to target vessels, typically guided by digital subtraction angiography (DSA)~\cite{cagnazzo2020endovascular} or X-ray fluoroscopy~\cite{hao2023casog}. However, challenges such as uneven contrast agent flow~\cite{meng2022weakly} or vascular occlusions~\cite{cagnazzo2020endovascular} may prevent some vessel branches from being opacified in intraoperative images. To ensure the safe deployment of instruments, the development of intelligent navigation systems is essential~\cite{abdelaziz2021x},~\cite{eagleton2022updates}. As a foundational element of navigation systems, real-time segmentation of vessel morphology is critically important~\cite{huang2024real},~\cite{zhu20233d}.

\input{Figure_Challenges}

However, accurately segmenting vessels from intraoperative images is non-trivial~\cite{ma2021self},~\cite{xia2019vessel},~\cite{huang2026vasomim}. As shown in Fig.~\ref{fig:challenges}, the primary challenges are threefold: \textbf{\textit{(i)}} To reduce radiation exposure for patients and physicians, interventions utilize low-power X-rays, resulting in \textit{low-signal-to-noise ratio (SNR) images}. \textbf{\textit{(ii)}} Complex vessel structures often include \textit{small and slender branches}, which are difficult to distinguish. \textbf{\textit{(iii)}} Non-target vessels, vessel-like objects (\textit{e.g.,} guidewires or catheters), and motion artifacts arising from patient physiological activities can cause \textit{significant interference}.

Early vessel segmentation methods rely on conventional image processing techniques. These methods begin by enhancing vessel features through image filters~\cite{frangi1998multiscale},~\cite{manniesing2006vessel},~\cite{cervantes2016segmentation}, followed by the application of region-growing~\cite{jiang2013region},~\cite{zeng2018automatic} or machine learning techniques~\cite{sangsefidi2018balancing},~\cite{wang2020tensor} to obtain segmentation results. However, these methods struggle to capture high-level semantic features crucial for successful segmentation, and their parameters are typically selected empirically~\cite{li2020cau}. Consequently, the robustness and generalization capabilities of these methods are inadequate, rendering them unsuitable for clinical applications.

In recent years, deep learning methods have played a dominant role in various vision tasks due to their powerful ability to learn high-level semantic features~\cite{minaee2021image},~\cite{chen2022recent},~\cite{xiao2026reason}. Specifically, in the medical image segmentation domain, UNet~\cite{ronneberger2015u} and its variants~\cite{zhou2019unet++},~\cite{schlemper2019attention},~\cite{huang2026mosformer} have gained widespread adoption, demonstrating remarkable success across different imaging modalities. Innovations based on UNet have aimed to further enhance vessel segmentation performance through various methods, including designing attention modules~\cite{li2020cau},~\cite{mou2021cs2}, exploiting full-resolution learning~\cite{liu2022full}, and integrating transformers~\cite{li2023global}. Despite these advancements, current methods have not fully addressed the challenges highlighted in Fig.~\ref{fig:challenges}, resulting in suboptimal performance. For example, CAU-net~\cite{li2020cau} utilizes a channel attention mechanism to learn channel-wise dependencies and mitigate interference, yet it struggles to accurately classify vessel structures in low-SNR images. Transformers~\cite{dosovitskiy2020image} offer advantages in capturing long-range vessel dependencies, which is beneficial for identifying vessel structures in low-SNR images~\cite{li2023global}. However, their image partitioning strategies~\cite{dosovitskiy2020image},~\cite{liu2021swin} may disrupt vessel continuity, adversely affecting precise identification of small or slender vessels.

Based on these observations, this paper proposes a novel \underline{\textbf{SP}}atial-frequency learning and graph-based channel \underline{\textbf{I}}nte\underline{\textbf{R}}acti\underline{\textbf{O}}n \underline{\textbf{Net}}work (\textbf{SPIRONet}) for vessel segmentation. \textbf{To address the challenges of \textit{(i) low-SNR} images and \textit{(ii) slender vessels}}, dual encoders are developed to learn from both the spatial and frequency domains. The CNN-based spatial encoder captures local features, such as vessel branches. In contrast, the Fourier-based frequency encoder models the global vessel structure and filters out noise. Considering that local spatial features and global frequency features are complementary and mutually enhancing~\cite{chen2024transunet},~\cite{guo2022cmt},~\cite{heidari2023hiformer},~\cite{kuang2024hybrid}, a cross-attention fusion module is designed to integrate these two types of features. This fusion enables SPIRONet to learn more discriminative vessel characteristics from \textit{low-SNR images} and accurately identify \textit{small or slender vessels}. \textbf{To mitigate \textit{(iii) structural interference},} a graph-based channel interaction module leverages graph neural networks (GNNs) to model explicit relationships between channel features, effectively suppressing task-irrelevant responses. These designs enable SPIRONet to effectively tackle the challenges outlined in Fig.~\ref{fig:challenges}.

Our contributions are summarized as follows:
\begin{itemize}
    \item To address low-SNR vessel appearance and difficulties in identifying small or slender branches, dual encoders are proposed to capture complementary spatial and frequency vessel representations. A cross-attention fusion module is further introduced to fuse local vessel details and global spectral cues.
    \item To suppress interference from non-target vessels and vessel-like structures, a graph-based channel interaction module is developed to model channel-wise correlations and enable effective information interaction, thereby filtering out task-irrelevant responses.
   \item SPIRONet achieves favorable performance compared with state-of-the-art alternatives on two in-house datasets (\textit{CADSA} and \textit{CAXF}) and three publicly available benchmarks (\textit{DCA1}, \textit{XCAD}, and \textit{ARCADE}) while achieving a real-time inference rate of 21 FPS\footnote{In intravascular intervention scenarios, the definition of ``real-time'' is $6-12$ FPS due to the low capture frequency of medical equipment such as X-ray systems~\cite{heidbuchel2014practical}.}.
\end{itemize}

The remainder of this paper is organized as follows: Section~\ref{sec:s2} briefly reviews related work. Section~\ref{sec:s3} describes SPIRONet in detail. Section~\ref{sec:s4} introduces datasets utilized in experiments and model configurations. Quantitative and qualitative experimental results are presented in Section~\ref{sec:s5}. Section~\ref{sec:s6} provides a detailed analysis of SPIRONet. Finally, Section~\ref{sec:s7} concludes this paper.

%% file: Figure_Challenges.tex
\begin{figure}[htbp]
	\centerline{\includegraphics{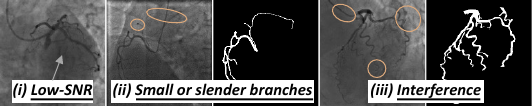}}
	\caption{Illustration of challenges in vessel segmentation. \textbf{\textit{(i)}} Low signal-to-noise ratio (SNR) images. \textbf{\textit{(ii)}} Small or slender vessel branches. \textbf{\textit{(iii)}} Non-target and motion artifact interference. X-ray fluoroscopy images and their corresponding ground truths are from the \textit{XCAD} dataset~\cite{ma2021self}.}
	\label{fig:challenges}
\end{figure}

%% file: sec2.tex
\subsection{Traditional Vessel Segmentation Approaches}Traditional vessel segmentation methods primarily rely on pixel intensity information. Among them, region-growing~\cite{adams1994seeded} is one of the most typical techniques. Jiang \textit{et al.}~\cite{jiang2013region} introduced an improved region-growing method that selects high-quality seeds based on spectral information. The ELEMENT~\cite{rodrigues2020element} framework was developed for multi-modal vessel segmentation. It integrates connectivity features with region-growing to identify potential vessel pixels and employs the Weka framework for segmentation by leveraging a comprehensive set of complementary features. In addition to region-growing methods, several other approaches have been explored. Dehkordi \textit{et al.}~\cite{taghizadeh2014local} proposed an active contour model that incorporates a local feature fitting energy for vessel segmentation. Similarly, Memari \textit{et al.}~\cite{memari2019retinal} adopted fuzzy C-means clustering to delineate coarse vessel structures, which are then refined using an integrated level set approach. Another innovative method, Tensor-cut~\cite{wang2020tensor}, conceptualizes each voxel as a second-order tensor and employs a graph-cut algorithm for final segmentation. Despite their effectiveness, these methods predominantly depend on manually designed features, requiring complex processing steps and facing scalability challenges~\cite{ma2021self},~\cite{kim2022diffusion}.

\subsection{Vessel Segmentation Based on Deep Learning}
With the tremendous advances in deep learning, researchers have designed various deep networks to improve vessel segmentation quality. UNet~\cite{ronneberger2015u} stands out for its encoder-decoder architecture, complemented by multi-scale skip connections, enabling the efficient capture of both low-level and high-level features. Building on UNet, models such as Attn-UNet~\cite{schlemper2019attention} and UNet++~\cite{zhou2019unet++} have been introduced to further enhance segmentation performance through the incorporation of attention mechanisms and the redesign of skip connections. For vessel-specific models, Gu \textit{et al.}~\cite{gu2019net} developed CE-Net, which employs a dense atrous convolution (DAC) and residual multi-kernel pooling (RMP) to simultaneously capture high-level features and preserve spatial vessel details. Similarly, CS$^{2}$-Net~\cite{mou2021cs2} integrates channel and spatial attention modules to improve feature representations. However, limited by the local receptive fields of convolutional neural networks (CNNs), these models cannot fully exploit global image context~\cite{dosovitskiy2020image}. In contrast, transformers exhibit a strong capacity for global context modeling~\cite{dosovitskiy2020image} and have been extensively applied in natural and medical image segmentation~\cite{chen2024transunet},~\cite{xie2021segformer}. TransUNet~\cite{chen2024transunet} is one of the pioneering methods that integrate transformers into medical image segmentation, utilizing transformers to encode CNN features for comprehensive global context modeling. UCTransNet~\cite{wang2022uctransnet} replaces the original skip connections with a channel transformer (CTrans) to reduce semantic gaps between shallower-level encoders and decoders. Additionally, Li \textit{et al.}~\cite{li2023global} proposed a global transformer and dual local attention network GT-DLA-dsHFF, which achieves deep-shallow hierarchical feature fusion to capture global and local vessel characteristics. Despite these advancements, existing deep learning-based approaches still have limitations in addressing the specific challenges (Fig.~\ref{fig:challenges}) of intraoperative vessel segmentation. CNN-based models are effective for extracting local vessel patterns, but their limited receptive fields may hinder the modeling of globally consistent vessel structures under low-SNR conditions. Transformer-based methods improve global context modeling through self-attention~\cite{dosovitskiy2020image,chen2024transunet}. However, their token- or patch-level representation may become less reliable when local regions are degraded by noise or low contrast, and may also weaken the continuity of tiny or slender vessels. Moreover, non-target vessels, guidewires, catheters, and motion artifacts often exhibit vessel-like appearances, making existing models prone to task-irrelevant responses and false-positive predictions. These limitations motivate the exploration of complementary spatial-frequency representations and graph-based channel interaction for more robust vessel segmentation.

\subsection{Learning from the Frequency Domain}
Fourier transform is a fundamental technique in conventional signal processing~\cite{gonzales1987digital}. Compared with purely spatial feature learning, frequency-domain analysis provides a complementary perspective for low-SNR images. In intraoperative images, weak vessel contrast and local noise fluctuations may corrupt pixel-wise or patch-wise responses, making spatial features less reliable. This problem can be more pronounced in purely spatial global self-attention, which models long-range dependencies based on token similarities~\cite{dosovitskiy2020image}. When local tokens are degraded by noise or low contrast, the estimated attention relationships may also become unreliable. In contrast, Fourier transform decomposes feature maps into global spectral components, which are less dependent on local spatial responses and can help capture more stable vessel-related structures under low-SNR conditions. Similar motivations have been explored in frequency-domain image denoising~\cite{cui2025adair} and medical image analysis~\cite{azad2021deep}. Recent studies have incorporated Fourier transform into neural networks through convolution~\cite{chi2020fast}, groupwise MLP layers~\cite{guibas2021adaptive}, element-wise multiplication with trainable parameters~\cite{rao2023gfnet}, and adaptive frequency filters~\cite{huang2023adaptive}. In medical image segmentation, FRCU-Net~\cite{azad2021deep} re-calibrates frequency components from Laplacian pyramids to produce more discriminative representations, while FDAM~\cite{huang2021medical} retains valuable frequency information of medical images. GFUNet~\cite{li2023global2} further replaces UNet's encoder with GFNet~\cite{rao2023gfnet} to exploit frequency features. However, these methods are not specifically designed for low-SNR vessel segmentation, and fixed or globally shared frequency operations may lack semantic adaptivity~\cite{huang2023adaptive}. To address this issue, this work separates frequency features into amplitude and phase components and employs lightweight convolutional layers to achieve semantically adaptive frequency modulation, aiming to enhance stable vessel-related cues while reducing unreliable local noise responses.

\input{Figure_Network}

\subsection{Channel Refinement Modules}
In high-level semantic feature maps, each channel often corresponds to a class-specific response, and different channel responses are often correlated with one another~\cite{fu2020scene},~\cite{li2022dual}. Many studies have been devoted to modeling channel correspondences to refine feature representations and eliminate task-irrelevant responses. Among these efforts, channel attention mechanisms (CAMs) have been extensively applied to various vision tasks~\cite{hu2019squeeze},~\cite{woo2018cbam},~\cite{wang2020eca}. Hu \textit{et al.}~\cite{hu2019squeeze} proposed a squeeze-and-excitation (SE) block, which recalibrates channel feature responses by explicitly modeling relationships between channels. ECA-Net~\cite{wang2020eca} introduces an efficient channel attention (ECA) module, enhancing the SE block to produce channel attention maps via 1D convolution without reducing dimensionality. Li \textit{et al.}~\cite{li2020cau} proposed CAU-net, which incorporates the SE block to capture vessel details and mitigate noise responses. Similarly, CAR-UNet~\cite{guo2021channel} integrates a channel attention double residual block (CADRB) to analyze channel statistics, while Mou \textit{et al.}~\cite{mou2021cs2} explored self-attention across channel dimensions to address long-range dependencies and refine channel features. However, channel correlations form a graph structure. The aforementioned methods allocate weights to each channel without explicitly formulating these topological relationships or directly modeling interactions among channel features~\cite{li2022dual}. As a result, they are less effective at enhancing vessel responses. A pioneering approach by~\cite{li2022dual} introduced a dynamic-channel graph convolutional network, mapping channels onto a topological space to enhance feature refinement on a graph. However, it is applied to feature maps with lower spatial resolution (\textit{i.e.,} at the bottom of the encoder), and irrelevant responses may be amplified again during decoding. In contrast, the proposed graph-based channel interaction module, activated post-decoding, aims to filter out irrelevant responses at higher resolution.

%% file: Figure_Network.tex
\begin{figure*}[t]
	\centering
	\centerline{\includegraphics{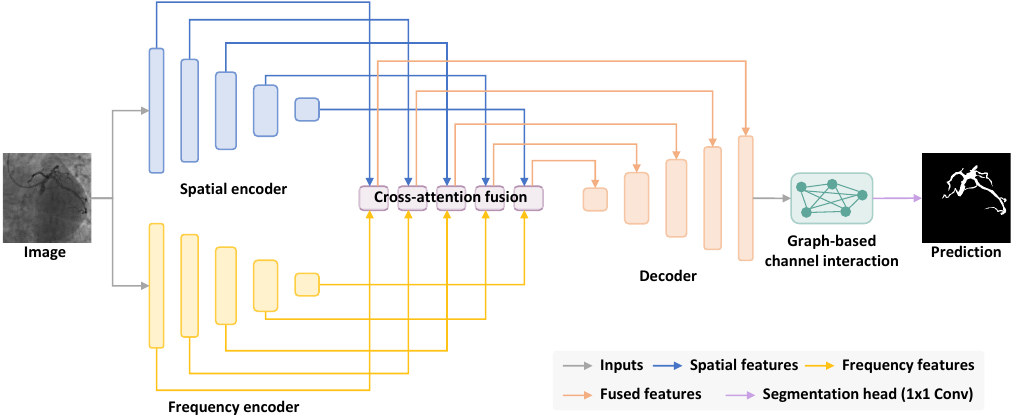}}
	\caption{SPIRONet adopts a spatial encoder and a frequency encoder to capture complementary spatial and frequency features. These two types of features are effectively fused by cross-attention fusion modules. Then, the fused features are fed into the decoder to recover the original resolution. After that, multi-channel features containing class-specific responses are refined by a graph-based channel interaction module based on GNNs. Finally, vessel predictions are obtained through a segmentation head.}
	\label{fig:network}
\end{figure*}

%% file: sec3.tex
The overall architecture of SPIRONet is illustrated in Fig.~\ref{fig:network}. It follows a U-shaped encoder-decoder architecture~\cite{ronneberger2015u} and utilizes two parallel encoders to learn local spatial features and global frequency features, respectively. A specially designed cross-attention fusion module is employed to mutually fuse spatial and frequency features. The fused features are then fed into the CNN decoder through skip connections. At the end of the decoder, a graph-based channel interaction module is deployed to refine channel features and filter out task-irrelevant responses. Finally, vessel predictions are generated through a segmentation head.

\subsection{Spatial-Frequency Representation Learning}

\textbf{Preliminaries.} 
Fourier transform plays a vital role in signal processing~\cite{gonzales1987digital} and is a key component in SPIRONet. Digital images can be regarded as 2D spatial signals, which can be transformed into the frequency domain via 2D discrete Fourier transform (DFT):
\begin{align}
    \bm{X}(u, v) =\sum_{h=0}^{H-1}\sum_{w=0}^{W-1}\bm{x}\left(h, w\right)e^{-j2\pi\left(\frac{uh}{H}+\frac{vw}{W}\right)} \label{eq:1}
\end{align}
where $\bm{x}\in \mathbb{R}^{1\times H\times W}$ is a single-channel image. $H$ and $W$ represent the height and width of $\bm{x}$, respectively. $u$ and $v$ are coordinates in the frequency domain. In practice, DFT is implemented using the fast Fourier transform (FFT)~\cite{cooley1965algorithm} with $\mathcal{O}(N\log N)$ complexity.

The amplitude $\bm{X}_{\rm Amp}$ and phase $\bm{X}_{\rm Pha}$ components are two important parts of $\bm{X}$. For a given $\bm{X}$, they can be derived as follows:
\begin{equation}\label{eq:amp_pha}
\begin{aligned}
    &\bm{X}_{\rm Amp}(u, v)
    = \sqrt{\left\{{\rm Re}\left[\bm{X}(u,v)\right]\right\}^2
    +\left\{{\rm Im}\left[\bm{X}(u,v)\right]\right\}^2} \\
    &\bm{X}_{\rm Pha}(u, v)
    = \arctan\left\{
    \frac{{\rm Im}\left[\bm{X}(u,v)\right]}
    {{\rm Re}\left[\bm{X}(u,v)\right]}
    \right\}
    \end{aligned}
\end{equation}

Similarly, given $\bm{X}_{\rm Amp}$ and $\bm{X}_{\rm Pha}$, the real and imaginary parts of $\bm{X}$ can be represented as:
\begin{equation}~\label{eq:real_imag}
    \begin{aligned}
        &{\rm Re}\left[\bm{X}(u,v)\right] = \bm{X}_{\rm Amp}(u, v)\cos\left\{\bm{X}_{\rm Pha}(u, v)\right\}\\
        &{\rm Im}\left[\bm{X}(u,v)\right] = \bm{X}_{\rm Amp}(u, v)\sin\left\{\bm{X}_{\rm Pha}(u, v)\right\}
    \end{aligned}
\end{equation}

\input{Figure_Encoder}

Eq.~\eqref{eq:1} shows that each frequency coefficient $\bm{X}(u,v)$ is computed from all spatial positions of the input image $\bm{x}$, indicating that the frequency-domain representation provides a global spectral view. This property is beneficial for low-SNR intraoperative images, where weak vessel contrast and local noise fluctuations may corrupt spatial-domain responses. By decomposing the image into spectral components, Fourier transform provides a complementary representation for capturing more stable vessel-related cues from noisy observations. The amplitude component $\bm{X}_{\rm Amp}$ reflects spectral intensity distributions related to contrast and texture, while the phase component $\bm{X}_{\rm Pha}$ preserves spatial structural information. Therefore, separately modeling amplitude and phase components enables the frequency encoder to enhance vessel-related spectral cues and alleviate the influence of unreliable local noise responses.

\textbf{Encoder Block.} 
Fig.~\ref{fig:encoder} illustrates the architectures of the proposed spatial and frequency encoder blocks. Given the inputs $\bm{f}_{\rm Spa}^i$ and $\bm{f}_{\rm Freq}^i$, the $i$-th spatial encoder block ${\rm Enc}_{\rm Spa}^i$ and frequency encoder block ${\rm Enc}_{\rm Freq}^i$ are used to extract local spatial features and frequency-domain vessel cues, respectively. Following previous medical image segmentation methods~\cite{chen2024transunet,kuang2024hybrid}, the spatial encoder adopts CNN to capture local vessel details, and its structure is similar to the residual module in ResNet~\cite{he2016deep}.

For the frequency encoder, let the input feature map be $\bm{f}_{\rm Freq}^i\in\mathbb{R}^{C_i\times H_i\times W_i}$. The 2D FFT is performed along the two spatial dimensions $(H_i,W_i)$ for each channel, while the channel dimension is preserved. In the implementation, we use real-valued FFT, resulting in a complex-valued spectrum $\bm{F}^i\in\mathbb{C}^{C_i\times H_i\times \tilde{W}_i}$, where $\tilde{W}_i=\lfloor W_i/2\rfloor+1$. The amplitude $\bm{F}_{\rm Amp}^i$ and phase $\bm{F}_{\rm Pha}^i$ are then obtained according to Eq.~\eqref{eq:amp_pha}.

Inspired by~\cite{wang2023spatial}, $\bm{F}_{\rm Amp}^i$ and $\bm{F}_{\rm Pha}^i$ are processed by two parallel convolutional paths:
\begin{equation}~\label{eq:amp_pha_modulation}
    \begin{aligned}
        &\widetilde{\bm{F}}_{\rm Amp}^i = \phi_{\rm Amp}^i\left(\bm{F}_{\rm Amp}^i\right) + \bm{F}_{\rm Amp}^i\\
        &\widetilde{\bm{F}}_{\rm Pha}^i = \phi_{\rm Pha}^i\left(\bm{F}_{\rm Pha}^i\right) + \bm{F}_{\rm Pha}^i
    \end{aligned}
\end{equation}
where $\phi_{\rm Amp}^i(\cdot)$ and $\phi_{\rm Pha}^i(\cdot)$ denote the amplitude and phase convolutional paths composed of $3\times3$ convolutions, respectively. Here, the $3\times3$ convolutions serve as learnable spectral modulation operations, rather than the direct source of global modeling. They model local interactions among neighboring frequency components in the amplitude and phase spectra.

The modulated amplitude and phase components are then integrated by reconstructing the complex spectrum. Specifically, the real and imaginary parts of the modulated spectrum $\widetilde{\bm{F}}^i$ are computed as:
\begin{equation}~\label{eq:freq_reconstruct}
    \begin{aligned}
        &{\rm Re}\left[\widetilde{\bm{F}}^i\right] = \widetilde{\bm{F}}_{\rm Amp}^i \cos\left(\widetilde{\bm{F}}_{\rm Pha}^i\right)\\
        &{\rm Im}\left[\widetilde{\bm{F}}^i\right] = \widetilde{\bm{F}}_{\rm Amp}^i \sin\left(\widetilde{\bm{F}}_{\rm Pha}^i\right)
    \end{aligned}
\end{equation}

After inverse FFT, the modulated spectral coefficients jointly reconstruct the spatial feature map. A residual connection and a $1\times1$ convolution are further used for feature refinement and channel adjustment:
\begin{align}~\label{eq:freq_encoder_output}
    \hat{\bm{f}}_{\rm Freq}^i =
    \bm{W}_{\rm c}^i
    \left(
    {\rm iFFT2}\left(\widetilde{\bm{F}}^i\right)
    + \bm{f}_{\rm Freq}^i
    \right)
\end{align}
where $\bm{W}_{\rm c}^i(\cdot)$ denotes the $1\times1$ convolution for channel adjustment. In this way, the amplitude and phase branches are not simply concatenated or added, but are integrated through complex-spectrum reconstruction followed by inverse FFT. Therefore, the frequency encoder provides global spectral cues that complement the local vessel details extracted by the spatial encoder. The frequency encoder block can be implemented in standard deep learning frameworks, such as PyTorch~\cite{paszke2019pytorch}, as summarized in Algorithm~\ref{algorithm:frequency_enc}.

\input{algorithm}
\input{Figure_Cross_Attn}

The outputs of the $i$-th encoder blocks are formulated as:\begin{equation}
    \begin{aligned}
        &\hat{\bm{f}}_{\rm Spa}^i = {\rm Enc}_{\rm Spa}^i\left(\bm{f}_{\rm Spa}^i\right) \\
        &\hat{\bm{f}}_{\rm Freq}^i = {\rm Enc}_{\rm Freq}^i\left(\bm{f}_{\rm Freq}^i\right)
    \end{aligned}
\end{equation}

Then, the outputs are downsampled by $2\times2$ maxpooling layers before being fed into the $(i+1)$-th encoder block.

\subsection{Cross-Attention Fusion}
As discussed in previous sections, the spatial encoder concentrates on learning local vessel features, while the frequency encoder captures long-range vessel dependencies, benefiting from the global spectral representation provided by the Fourier transform. Recent research indicates that local and global features are complementary, providing mutual guidance for learning more robust features~\cite{guo2022cmt},~\cite{kuang2024hybrid}. To effectively integrate local spatial features and global frequency features, a cross-attention fusion module is proposed, as depicted in Fig. ~\ref{fig:cross_attn}.

The inputs to the $i$-th cross-attention fusion module are features ($\hat{\bm{f}}_{\rm Spa}^i, \hat{\bm{f}}_{\rm Freq}^i\in \mathbb{R}^{C_i\times H_i \times W_i}$) extracted by the $i$-th spatial and frequency encoder blocks, where $C_i$, $H_i$, and $W_i$ represent the channel number, height, and width of feature maps. First, these two features are projected into the embedding space through different “Conv-BN-ReLU” layers $\bm{W}^i(\cdot)$ to generate the spatial query $\bm{Q}^i_{\rm Spa}$ and key $\bm{K}^i_{\rm Spa}$, as well as the frequency query $\bm{Q}^i_{\rm Freq}$ and key $\bm{K}^i_{\rm Freq}$. The mixed value is the projection of the concatenated feature ${\rm Concat}\left(\hat{\bm{f}}_{\rm Spa}^i, \hat{\bm{f}}_{\rm Freq}^i\right)\in \mathbb{R}^{2C_i\times H_i\times W_i}$.

To alleviate the high computational complexity caused by matrix multiplications, pyramid pooling modules (PPM)~\cite{zhu2019asymmetric} are adopted to sample the key and value features. Specifically, the embedding dimension is set to $d_0=C_i/8$. Given a projected feature map $\bm{z}\in\mathbb{R}^{d_0\times H_i\times W_i}$, PPM applies adaptive average pooling with output scales $\mathcal{S}=\{1,3,6,8\}$. For each scale $s\in\mathcal{S}$, the pooled feature map has the size $d_0\times s\times s$ and is flattened into $s^2$ tokens. The tokens from all scales are concatenated along the spatial dimension:
\begin{equation}
    {\rm PPM}(\bm{z}) =
    {\rm Concat}_{s\in\mathcal{S}}
    \left[
    {\rm Flatten}
    \left(
    {\rm AdaAvgPool}_{s\times s}(\bm{z})
    \right)
    \right]
\end{equation}Thus, ${\rm PPM}(\bm{z})\in\mathbb{R}^{N\times d_0}$, where $N=\sum_{s\in\mathcal{S}}s^2=1^2+3^2+6^2+8^2=110$. It should be noted that PPM is only applied to the key and value branches, while the query branches retain the original spatial resolution. Therefore, the cross-attention output is still computed for each spatial location, which helps preserve dense responses for small or slender vessels.

This process is formulated as follows:
\begin{align}
    &\bm{Q}^i_{\rm Spa} = \bm{W}^i_{\rm SQ}\left(\hat{\bm{f}}_{\rm Spa}^i\right), \bm{Q}^i_{\rm Freq} = \bm{W}^i_{\rm FQ}\left(\hat{\bm{f}}_{\rm Freq}^i\right) \\
    &\resizebox{.9\hsize}{!}{$\bm{K}^i_{\rm Spa} = {\rm PPM}\left\{\bm{W}^i_{\rm SK}\left(\hat{\bm{f}}_{\rm Spa}^i\right)\right\}, \bm{K}^i_{\rm Freq} = {\rm PPM}\left\{\bm{W}^i_{\rm FK}\left(\hat{\bm{f}}_{\rm Freq}^i\right)\right\}$} \\
    & \bm{V}^i = {\rm PPM}\left\{\bm{W}^i_{\rm V}\left[{\rm Concat}\left(\hat{\bm{f}}_{\rm Spa}^i, \hat{\bm{f}}_{\rm Freq}^i\right)\right]\right\}
\end{align}
where $\bm{Q}^i_{\rm Spa}, \bm{Q}^i_{\rm Freq}\in \mathbb{R}^{\left(H_i\cdot W_i\right)\times d_0}$, $\bm{K}^i_{\rm Spa}, \bm{K}^i_{\rm Freq}\in \mathbb{R}^{N\times d_0}$, and $\bm{V}^i\in \mathbb{R}^{N\times d_0}$. $d_0$ is the dimension of the embedding space and $N$ is the number of tokens sampled by PPM, $N\ll H_i\cdot W_i$. Thus, CA can be calculated as follows:
\begin{align}
    \resizebox{.9\hsize}{!}{${\rm CA}\left(\bm{Q}^i, \bm{K}^i, \bm{V}^i\right) = {\rm Softmax}\left(\frac{\bm{Q}^i_{\rm Spa}\bm{K}^{iT}_{\rm Freq}+\bm{Q}^i_{\rm Freq}\bm{K}^{iT}_{\rm Spa}}{\sqrt{d_0}}\right)\bm{V}^i$}
\end{align}

\subsection{Graph-based Channel Interaction}
As visualized in Fig.~\ref{fig:tci_vis}, which overlays channel features on their corresponding images, different channels exhibit highly varied activation patterns. While some share similar or complementary responses (row 1), others capture distinctly different information (row 2). This observation suggests a potential way to enhance vessel-specific features for segmentation. Prior methods typically attempt this by simply re-weighting channels using predefined attention modules. However, these approaches fail to facilitate direct interaction between channels, thereby limiting the fusion of different features.

\input{Figure_TCI_Vis}
\input{Figure_TCI}

Inspired by~\cite{li2022dual}, this study overcomes this limitation by formulating the channel relationships as a graph. Building upon this, a graph-based channel interaction (GCI) module is introduced, as illustrated in Fig.~\ref{fig:graph}. It enables direct information exchange among channel nodes, thereby more effectively fusing features and suppressing noise. Graph neural networks (GNNs) are widely utilized to learn features of graph-structured data. Given a graph $\mathcal{G}=\left(V, E\right)$ and its adjacency matrix $\bm{A}$, the output of a GNN can be formulated as follows~\cite{kipf2016semi}:
\begin{align}
    &\bm{H}_{\rm out} = \sigma\left(\tilde{\bm{L}}\bm{H}_{\rm in}\bm{\Theta}\right) \label{eq:graph}\\
    &\tilde{\bm{L}} = \tilde{\bm{D}}^{-\frac{1}{2}}\tilde{\bm{A}}\tilde{\bm{D}}^{-\frac{1}{2}}
\end{align}
where $\sigma\left(\cdot\right)$ is a non-linear activation function and $\bm{\Theta}$ denotes trainable parameters. $\tilde{\bm{L}}$ is the Laplacian matrix, $\tilde{\bm{A}} = \bm{A} + \bm{I}$, and $\tilde{\bm{D}}_{ii} = \sum_{j}\tilde{\bm{A}}_{ij}$.

The input to the graph-based channel interaction module is $\bm{H}_{\rm in} = \bm{f}\in \mathbb{R}^{C\times (H\cdot W)/4^2}$, where the outputs of the decoder $\bm{f}_{\rm in}\in \mathbb{R}^{C\times H\times W}$ are downsampled via a $4\times 4$ maxpooling layer. To learn data-dependent graph structures, this work follows~\cite{li2020spatial} and uses an improved Laplacian matrix:
\begin{align}
    \tilde{\bm{L}} = \bm{I} - \tilde{\bm{D}}^{-\frac{1}{2}}\tilde{\bm{A}}\tilde{\bm{D}}^{-\frac{1}{2}}
\end{align}
where $\tilde{\bm{A}}, \tilde{\bm{D}} \in \mathbb{R}^{C\times C}$. $C$ denotes the number of channels of input feature maps $\bm{f}$.

To formulate relationships among different channels, the cosine similarity between channels is calculated as:
\begin{align}
    \tilde{a}_{ij} = \frac{\bm{W}\left(\bm{f}\right)_i^T\bm{W}\left(\bm{f}\right)_j}{||\bm{W}\left(\bm{f}\right)_i||_2 ||\bm{W}\left(\bm{f}\right)_j||_2}
\end{align}
where $\bm{W}(\cdot)$ is an embedding layer consisting of ``Conv-ReLU" and $\tilde{a}_{ij} \in \tilde{\bm{A}}$. Thus Eq.~\eqref{eq:graph} can be written as follows:
\begin{align}
    \bm{H}_{\rm out} = \sigma\left\{\tilde{\bm{L}}{\rm Down}_{/4}\left(\bm{f}_{\rm in}\right)\bm{\Theta}\right\}
\end{align}
where $\sigma$ is ReLU and ${\rm Down}_{/4}$ indicates the downsample operator. Overall, the outputs of GCI can be formulated as:
\begin{align}
    \bm{f}_{\rm out} = {\rm Up}_{\times4}\left\{\sigma\left[\tilde{\bm{L}}{\rm Down}_{/4}\left(\bm{f}_{\rm in}\right)\bm{\Theta}\right]\right\} + \bm{f}_{\rm in}
\end{align}
where ${\rm Up}_{\times4}$ indicates the upsample operator. Specifically, downsample and upsample operators are implemented via maxpooling and transposed convolution, respectively.

%% file: Figure_Encoder.tex
\begin{figure}[htbp]
	\centerline{\includegraphics{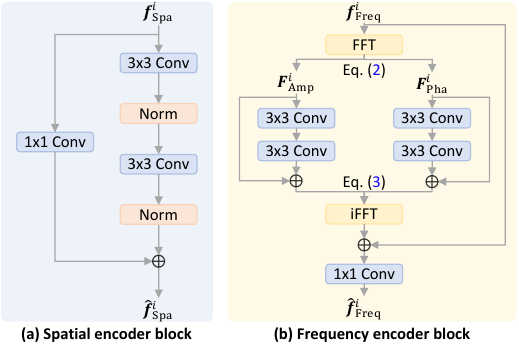}}
	\caption{The architecture of encoder blocks. (a) Spatial encoder block. (b) Frequency encoder block. $\oplus$ denotes element-wise addition.}
	\label{fig:encoder}
\end{figure}

%% file: algorithm.tex
\begin{algorithm}[htbp]
\caption{Pseudo-code of the frequency encoder block in a PyTorch-like style.}
\label{algorithm:frequency_enc}
\begin{lstlisting}[language=python, morekeywords={rfft2,irfft,conv_amp,conv_pha,conv_channel,angle,cos,sin}]
# x: input feature maps
# conv_amp, conv_pha: amplitude/phase convolutional layers
# conv_channel: channel adjustment

# fast Fourier transform (FFT) Eq. (1)
X = rfft2(x)
# decompose X into amplitude and phase components Eq. (2)
amp = abs(X)
pha = angle(X)
# parallel paths
amp_fuse = conv_amp(amp) + amp
pha_fuse = conv_pha(pha) + pha
# get real and imaginary parts Eq. (3)
real = amp_fuse * cos(pha_fuse)
imag = amp_fuse * sin(pha_fuse)
# inverse fast Fourier transform (iFFT)
X_ = complex(real, imag)
x_ = irfft2(X_)
# final outputs
out = x_ + x
out = conv_channel(out)
\end{lstlisting}
\end{algorithm}

%% file: Figure_Cross_Attn.tex
\begin{figure}[htbp]
	\centerline{\includegraphics{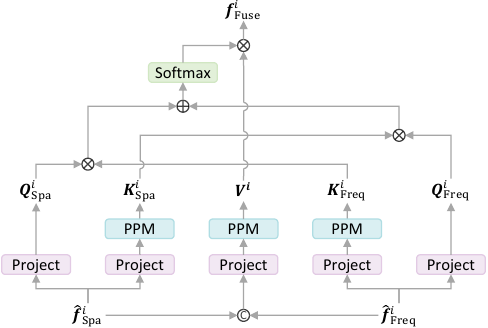}}
	\caption{Cross-attention module. $\otimes$, $\oplus$, and $\textcircled{c}$ denote matrix multiplication, element-wise addition, and channel-wise concatenation.}
	\label{fig:cross_attn}
\end{figure}

%% file: Figure_TCI_Vis.tex
\begin{figure}[htbp]
    \centerline{\includegraphics[width=\linewidth]{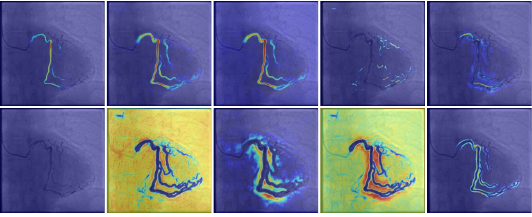}}
    \caption{Visualization of several channel features. The features overlay the original image, where red indicates a strong response and blue indicates a weak response. Some channels exhibit similar or complementary response patterns (row 1), while others capture highly distinct information (row 2).
    }
    \label{fig:tci_vis}
\end{figure}

%% file: Figure_TCI.tex
\begin{figure}[htbp]
    \centerline{\includegraphics{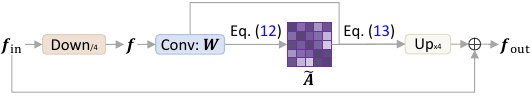}}
    \caption{{Graph-based channel interaction module.} $\oplus$ represents element-wise addition.}
    \label{fig:graph}
\end{figure}

%% file: sec4.tex
\subsection{Datasets}
Five challenging datasets are used in this paper, comprising two in-house datasets and three publicly available datasets. To prevent data leakage, all in-house datasets are partitioned at the patient level rather than the image level. For the public datasets, we follow the official splits.

\textit{\textbf{Chronic Artery Digital Subtraction Angiography Dataset (CADSA)}}. The \textit{CADSA} dataset is derived from our collaborating hospital. It contains 275 images from 21 patient DSA sequences, with an original resolution of $750\times 750$. Chronic arteries are annotated by experienced physicians using ITK-SNAP~\cite{yushkevich2006user}. 197 images from DSA sequences of 15 patients are randomly selected for training, and the remaining 78 images from 6 patients’ DSA sequences are used for testing.

\textit{\textbf{Coronary Arteries X-ray Fluoroscopy Dataset (CAXF)}}. This dataset comprises 538 images from 36 X-ray fluoroscopy sequences, with a resolution of $512\times 512$. Specifically, 412 images are from 7 sequences, and 126 images are randomly selected from 29 sequences. Following previous settings~\cite{li2020cau}, 337 images from 24 sequences are utilized for training, and the remaining 201 images from 12 sequences are selected for testing. No patient overlap exists between the training and testing sets.

\textit{\textbf{DCA1}}~\cite{cervantes2019automatic}. This dataset is provided by the Mexican Social Security Institute, UMAE T1-Le\'on. It includes 134 $300\times 300$ X-ray coronary angiograms along with corresponding ground truths annotated by an expert cardiologist. Following~\cite{liu2022full}, this dataset is split into 100 training images and 34 testing images.

\textit{\textbf{XCAD}}~\cite{ma2021self}. The \textit{XCAD} dataset contains 1,747 coronary angiograms obtained by a General Electric Innova IGS 520 system. Each image has a resolution of $512\times 512$. It is important to note that only 126 images in this dataset have vessel annotations. The training set comprises 100 images, while the remaining 26 images constitute the testing set.

\input{Table_Dataset}

\textbf{\textit{ARCADE}}~\cite{popov2024dataset}. This dataset consists of two subsets: one for coronary vessel segmentation and another for stenosis detection. Each subset includes 1,500 images at a $512\times512$ resolution, partitioned into training (1,000), validation (200), and testing (300) sets. This study focuses on the vessel segmentation subset.

\subsection{Evaluation Metrics}
To thoroughly evaluate the proposed method and baselines, four evaluation metrics are selected, including sensitivity (Sen.), F1-score (F1), Intersection over Union (IoU), and Matthews correlation coefficient (MCC).
\begin{align}
    &\text{Sen.} = \frac{\text{TP}}{\text{TP} + \text{FN}} \\
    &\text{F1} = \frac{2 \times \text{TP}}{2 \times \text{TP} + \text{FP} + \text{FN}} \\
    &\text{IoU} = \frac{\text{TP}}{\text{TP} + \text{FP} + \text{FN}} \\
    &\resizebox{.9\hsize}{!}{$\text{\scriptsize MCC} = \frac{\text{TP} \times \text{TN} - \text{FP} \times \text{FN}}{\sqrt{(\text{TP} + \text{FP}) \times (\text{FP} + \text{FN}) \times (\text{TN} + \text{FP}) \times (\text{TN} + \text{FN})}}$}
\end{align}
where TP, TN, FP, and FN represent true-positive, true-negative, false-positive, and false-negative pixels in segmentation results, respectively.

\subsection{Implementation Details}
All experiments are implemented based on PyTorch 1.12.0~\cite{paszke2019pytorch}, Python 3.8, and Ubuntu 18.04. For the \textit{CADSA}, \textit{CAXF}, \textit{DCA1}, and \textit{XCAD} datasets, training is conducted on a single NVIDIA GeForce RTX 3090 GPU (24GB). The larger \textit{ARCADE} dataset is trained on a single NVIDIA H20 GPU (96GB). Common training settings include data augmentation with random flipping and rotations within $\left[-20^{\circ}, 20^{\circ}\right]$. Binary cross-entropy (BCE) loss is adopted as the training objective. Input images are resized to $512\times512$, except for the \textit{DCA1} dataset, whose images are $300\times300$. The learning rate is decayed using a polynomial annealing policy: $lr \leftarrow lr_{\rm init} * \left(1 - \frac{\rm epoch}{\rm epoch_{\rm total}}\right)^{0.9}$. To ensure robust results, all experiments are repeated three times using different random seeds. The dataset-specific training configurations are detailed in Table~\ref{table:dataset}.

%% file: Table_Dataset.tex
\begin{table}[htbp]
\caption{
Detailed training configurations of each dataset.
}
\label{table:dataset}
\centering
\renewcommand\arraystretch{1.2}
\resizebox{\linewidth}{!}{
\begin{tabular}{llccc}
\toprule
Dataset & Optimizer & Batch Size & ${lr_{\rm init}}$ & ${\rm epoch_{\rm total}}$ \\ \midrule
\textit{CADSA} & \multirow{4}{*}{SGD} & \multirow{4}{*}{$4$} & $3.0e^{-2}$ & $200$ \\
\textit{CAXF}~\cite{li2020cau} & & & $3.0e^{-2}$ & $800$ \\
\textit{DCA1}~\cite{cervantes2019automatic} & & & $4.5e^{-2}$ & $600$ \\
\textit{XCAD}~\cite{ma2021self} & & & $6.0e^{-2}$ & $700$ \\
\midrule
\textit{ARCADE}~\cite{popov2024dataset} & Adam & $24$ & $3.0e^{-4}$ & $100$ \\ \bottomrule
\end{tabular}
}
\end{table}

%% file: sec5.tex
In this section, extensive experiments are conducted to answer the following questions:
\begin{itemize} 
    \item \textit{\textbf{Q1}}: Does the proposed SPIRONet yield better vessel segmentation performance than state-of-the-arts?
    \item \textit{\textbf{Q2}}: Do the proposed modules improve vessel segmentation performance?
    \item \textit{\textbf{Q3}}: Does SPIRONet achieve a desirable trade-off between computational efficiency and segmentation performance?
\end{itemize}

\subsection{Comparisons with State-of-the-Arts (Q1)}
SPIRONet is evaluated against several state-of-the-art (SOTA) models across five datasets. For a fair comparison, all baseline models are trained using their official implementations under the same training configurations as SPIRONet.

\input{Table_SOTA_CADSA_CAXF}
\input{Table_SOTA_DCA1_XCAD}

\textbf{In-house Datasets.} Table~\ref{table:cadsa_caxf} summarizes quantitative results on two in-house datasets, \textit{CADSA} and \textit{CAXF}~\cite{li2020cau}. Notably, the proposed SPIRONet achieves $81.20\%$ sensitivity, $80.10\%$ F1, $68.27\%$ IoU, and $80.61\%$ MCC on the \textit{CADSA} dataset and $90.80\%$ sensitivity, $90.32\%$ F1, $82.48\%$ IoU, and $89.94\%$ MCC on the \textit{CAXF} dataset, outperforming the compared SOTA models. On the \textit{CADSA} dataset, it is worth noting the $5.11\%$ sensitivity improvement of SPIRONet over CAU-net~\cite{li2020cau} (the second-best method). Higher sensitivity indicates that models are more capable of extracting thin vessels and vessel boundaries~\cite{liu2022full}, demonstrating that SPIRONet has a strong ability to learn discriminative vessel features.

\textbf{Publicly Available Benchmarks.} Table~\ref{table:dca1_xcad} presents comparison results on two publicly available benchmarks, \textit{DCA1}~\cite{cervantes2019automatic} and \textit{XCAD}~\cite{ma2021self}. On the \textit{DCA1} dataset, SPIRONet achieves the best F1 and IoU, while showing sensitivity and MCC comparable to those of SOTA models. SPIRONet obtains the best results in all metrics on the \textit{XCAD} dataset. To further validate its scalability and robustness, experiments are conducted on the larger \textit{ARCADE} dataset. The results in Table~\ref{table:arcade} confirm that SPIRONet consistently surpasses all baselines by a clear margin. Collectively, these results across three public datasets demonstrate the effectiveness and generalization potential of SPIRONet for vessel segmentation.

\input{Table_TTest}
\input{Figure_SOTA_Vis}
\input{Table_Ablation}
\input{Table_SOTA_ARCADE}
\input{Table_TTest_All_Datasets}

\textbf{Statistical Analysis.} To further verify whether the performance gains of SPIRONet are statistically reliable, we first conduct paired image-level statistical analysis on the \textit{ARCADE} dataset. We select \textit{ARCADE} for this detailed analysis because it has a relatively large testing set among the evaluated datasets, containing 300 testing images, which provides a more representative basis for image-level statistical testing. For each method, the per-image segmentation metrics are first averaged over three random seeds. For each metric, we perform a two-sided Wilcoxon signed-rank test. In addition, the $95\%$ confidence interval of the mean improvement is estimated using bootstrap resampling over testing images. As shown in Table~\ref{tab:arcade_statistical_analysis}, SPIRONet achieves statistically significant improvements over all compared methods on the \textit{ARCADE} dataset. To further examine the statistical reliability of SPIRONet across different datasets, we additionally conduct paired image-level statistical analysis between SPIRONet and the strongest competing method on each dataset, as shown in Table~\ref{table:wilcoxon_all_datasets}. For each dataset, the strongest competing method is selected according to the average IoU. The results show that SPIRONet achieves statistically significant IoU improvements on \textit{CAXF}, \textit{XCAD}, and \textit{ARCADE}. On \textit{DCA1}, the improvement is close to the conventional significance threshold, while on \textit{CADSA}, the improvement is not statistically significant. Overall, these results provide a more comprehensive statistical assessment across datasets, suggesting that SPIRONet consistently improves the average IoU over the strongest competing methods, whereas the statistical significance varies across datasets.

\textbf{Visualization Results.} Segmentation results of several models are further visualized to provide qualitative comparisons, including UNet~\cite{ronneberger2015u}, UNet++~\cite{zhou2019unet++}, CE-Net~\cite{gu2019net}, CAU-net~\cite{li2020cau}, CS$^2$-Net~\cite{mou2021cs2}, DE-DCGCN-EE~\cite{li2022dual}, and SPIRONet. These visualizations intuitively show how SPIRONet addresses the challenges presented in Fig.~\ref{fig:challenges}: \textbf{\textit{(i)}} Uneven contrast-agent flow and low-power X-ray imaging cause the images in rows 1 and 3 to exhibit \textit{low-SNR}. The proposed model demonstrates a strong ability to locate challenging vessel regions, as highlighted in red boxes. \textbf{\textit{(ii)}} The \textit{slender} vessels in rows 2 and 5, highlighted in red boxes, are often incompletely segmented by the baselines. By fully exploring local spatial and global frequency features, SPIRONet achieves more precise segmentation results. \textbf{\textit{(iii)}} Non-target vessels highlighted in red boxes in row 4 cause \textit{interference} for segmentation, resulting in some false-positive predictions by baselines. Benefiting from the proposed graph-based channel interaction module, SPIRONet successfully suppresses irrelevant interference and avoids false-positive predictions.

\subsection{Ablation Studies (Q2)} \label{sec: ablation}
In this section, experiments are conducted on the \textit{XCAD} and \textit{ARCADE} datasets to evaluate the effectiveness of key components in SPIRONet. The quantitative results are reported in Table~\ref{table:ablation}. For variants lacking the cross-attention module (\textit{e.g.,} Variant III and VII), spatial and frequency features are fused via simple element-wise addition. The main findings are as follows: \textbf{\textit{(i)}} Learning from both spatial and frequency domains is crucial for precise vessel segmentation. Models combining both spatial and frequency encoders (Variant III) consistently outperform single-encoder variants (I and II) on both datasets. \textbf{\textit{(ii)}} On the smaller \textit{XCAD} dataset, the spatial-only model (Variant I) is superior to the frequency-only model (Variant II).\input{Table_Param}This suggests that the frequency encoder, similar to vision transformers~\cite{dosovitskiy2020image}, may require a relatively large amount of data to effectively learn long-range dependencies. This observation is consistent with the results on the larger \textit{ARCADE} dataset, where the performance gap narrows significantly, with Variant II achieving results comparable to Variant I. This indicates the frequency encoder's potential is better realized with more training data. \textbf{\textit{(iii)}} The cross-attention (CA) module provides a more effective fusion mechanism than element-wise addition. Comparing Variant IV against Variant III demonstrates a clear performance uplift. The improvement is particularly notable on the \textit{ARCADE} dataset, where IoU increases from $59.88\%$ to $62.78\%$, validating the CA module's ability to effectively fuse spatial and frequency features. \textbf{\textit{(iv)}} Incorporating the graph-based channel interaction (GCI) module into the network (Variant IV) yields consistent performance gains across all metrics. Notably, it improves sensitivity by $+0.91\%$ on \textit{XCAD} and a substantial $+4.19\%$ on \textit{ARCADE}. This highlights the GCI module's effectiveness in filtering task-irrelevant responses and enhancing the final segmentation quality.

\input{Table_FrequencyFilter}

\subsection{Model Complexity (Q3)}
The computational cost of SPIRONet is analyzed to evaluate its practical feasibility. Table~\ref{table:complexity} presents a detailed comparison of model parameters, floating-point operations (FLOPs), and inference rates, calculated on a single NVIDIA GeForce RTX 3090 GPU using the thop library\footnote{\href{https://pypi.org/project/thop/}{\tt \color{blue}https://pypi.org/project/thop/}.}. A key finding is that SPIRONet meets real-time requirements for clinical deployment. Its inference rate of approximately 21 FPS is nearly double the upper end of the $6-12$ FPS range considered real-time for interventional procedures~\cite{heidbuchel2014practical}.

Furthermore, when compared against other models with high performance, SPIRONet demonstrates a favorable efficiency profile. It maintains a moderate parameter count (16.98M), which is significantly lower than other competitive architectures like TransUNet (93.23M) and AttnUNet (57.16M). Its computational cost (200.90G FLOPs) is also competitive. While lightweight models like CAU-net~\cite{li2020cau} are faster, they achieve relatively lower segmentation results, particularly on challenging datasets like \textit{ARCADE}.

Therefore, SPIRONet strikes a favorable balance: it delivers a significant performance improvement over simpler models and offers greater efficiency than other complex, high-performing alternatives. This indicates a favorable trade-off, positioning SPIRONet as a powerful yet practical solution for clinical application.

%% file: Table_SOTA_CADSA_CAXF.tex
\begin{table*}[t]
\caption{Comparison with state-of-the-arts on the \textit{CADSA} and \textit{CAXF} datasets. The best results are highlighted in \sethlcolor{blue!30}\hl{blue} and the second-best results are highlighted in \sethlcolor{red!30}\hl{red}. ``mean $\pm$ std'' is reported over three random seeds.}
\label{table:cadsa_caxf}
\centering
\renewcommand\arraystretch{1.2}
\resizebox{\linewidth}{!}{
\begin{tabular}{lcccccccc}
\toprule
\multirow{2}{*}{Model} & \multicolumn{4}{c|}{\textit{CADSA}} & \multicolumn{4}{c}{\textit{CAXF}} \\ \cmidrule{2-9} 
& Sen. (\%) $\uparrow$ & F1 (\%) $\uparrow$ & IoU (\%) $\uparrow$ & \multicolumn{1}{l|}{MCC (\%) $\uparrow$} & Sen. (\%) $\uparrow$  & F1 (\%) $\uparrow$  & IoU (\%) $\uparrow$  & MCC (\%) $\uparrow$ \\ \midrule
UNet~\cite{ronneberger2015u}~{\tiny \color{gray}[MICCAI'15]} & $72.11${\tiny $\pm 4.05$} & $76.58${\tiny $\pm 2.90$} & $64.60${\tiny $\pm 3.20$} & \multicolumn{1}{l|}{$77.58${\tiny $\pm 2.59$}} & $89.22${\tiny $\pm 0.42$} & $89.90${\tiny $\pm 0.18$} & $81.80${\tiny $\pm 0.29$} & $89.50${\tiny $\pm 0.18$} \\
UNet++~\cite{zhou2019unet++}~{\tiny \color{gray}[TMI'19]} & $74.59${\tiny $\pm 4.46$} & $78.04${\tiny $\pm 2.26$} & $66.20${\tiny $\pm 2.49$} & \multicolumn{1}{l|}{$78.88${\tiny $\pm 1.86$}} & $89.53${\tiny $\pm 0.28$} & \cellcolor{red!30}$90.01${\tiny $\pm 0.18$} & \cellcolor{red!30}$81.96${\tiny $\pm 0.28$} & \cellcolor{red!30}$89.59${\tiny $\pm 0.18$} \\
AttnUNet~\cite{schlemper2019attention}~{\tiny \color{gray}[MedIA'19]} & $73.27${\tiny $\pm 5.70$} & $76.58${\tiny $\pm 2.81$} & $64.34${\tiny $\pm 3.14$} & \multicolumn{1}{l|}{$77.79${\tiny $\pm 2.20$}} & $89.13${\tiny $\pm 0.15$} & $89.76${\tiny $\pm 0.09$} & $81.58${\tiny $\pm 0.15$} & $89.34${\tiny $\pm 0.09$} \\
CE-Net~\cite{gu2019net}~{\tiny \color{gray}[TMI'19]} & $75.61${\tiny $\pm 2.69$} & $77.80${\tiny $\pm 0.93$} & $65.20${\tiny $\pm 1.20$} & \multicolumn{1}{l|}{$78.69${\tiny $\pm 0.84$}} & $89.74${\tiny $\pm 0.37$} & $89.93${\tiny $\pm 0.13$} & $81.82${\tiny $\pm 0.22$} & $89.52${\tiny $\pm 0.14$} \\
CAU-net~\cite{li2020cau}~{\tiny \color{gray}[ICONIP'20]} & \cellcolor{red!30}$76.09${\tiny $\pm 0.17$} & \cellcolor{red!30}$79.05${\tiny $\pm 1.43$} & \cellcolor{red!30}$67.40${\tiny $\pm 1.57$} & \multicolumn{1}{l|}{\cellcolor{red!30}$79.88${\tiny $\pm 1.29$}} & $89.31${\tiny $\pm 0.27$} & $89.85${\tiny $\pm 0.16$} & $81.70${\tiny $\pm 0.23$} & $89.43${\tiny $\pm 0.15$} \\
CS$^2$-Net~\cite{mou2021cs2}~{\tiny \color{gray}[MedIA'21]} & $66.12${\tiny $\pm 5.08$} & $73.10${\tiny $\pm 3.80$} & $60.29${\tiny $\pm 4.02$} & \multicolumn{1}{l|}{$74.95${\tiny $\pm 3.08$}} & $89.84${\tiny $\pm 0.10$} & $89.84${\tiny $\pm 0.03$} & $81.67${\tiny $\pm 0.07$} & $89.41${\tiny $\pm 0.04$} \\
FR-UNet~\cite{liu2022full}~{\tiny \color{gray}[JBHI'22]} & $59.94${\tiny $\pm 0.77$} & $62.70${\tiny $\pm 0.42$} & $50.72${\tiny $\pm 0.31$} & \multicolumn{1}{l|}{$63.55${\tiny $\pm 0.36$}} & $90.10${\tiny $\pm 0.78$} & $89.17${\tiny $\pm 0.14$} & $80.56${\tiny $\pm 0.21$} & $88.71${\tiny $\pm 0.14$} \\
DE-DCGCN-EE~\cite{li2022dual}~{\tiny \color{gray}[TMI'22]} & $66.42${\tiny $\pm 1.90$} & $70.57${\tiny $\pm 0.55$} & $57.14${\tiny $\pm 0.74$} & \multicolumn{1}{l|}{$72.16${\tiny $\pm 0.54$}} & $89.29${\tiny $\pm 0.28$} & $88.91${\tiny $\pm 0.07$} & $80.16${\tiny $\pm 0.10$} & $88.45${\tiny $\pm 0.08$} \\
GT-DLA-dsHFF~\cite{li2023global}~{\tiny \color{gray}[TCYB'23]} & $70.14${\tiny $\pm 7.32$} & $71.32${\tiny $\pm 3.09$} & $57.83${\tiny $\pm 2.93$} & \multicolumn{1}{l|}{$72.67${\tiny $\pm 2.57$}} & $89.94${\tiny $\pm 0.07$} & $89.94${\tiny $\pm 0.13$} & $81.82${\tiny $\pm 0.20$} & $89.52${\tiny $\pm 0.13$} \\ 
TransUNet~\cite{chen2024transunet}~{\tiny \color{gray}[MedIA'24]} & $74.59${\tiny $\pm 1.69$} & $75.80${\tiny $\pm 2.80$} & $63.60${\tiny $\pm 2.91$} & \multicolumn{1}{l|}{$76.86${\tiny $\pm 2.48$}} & \cellcolor{red!30}$90.16${\tiny $\pm 0.18$} & $89.89${\tiny $\pm 0.09$} & $81.75${\tiny $\pm 0.14$} & $89.47${\tiny $\pm 0.09$} \\
\midrule
SPIRONet~{\tiny \color{gray}\textbf{[Ours]}} & \cellcolor{blue!30}$81.20${\tiny $\pm 2.72$} & \cellcolor{blue!30}$80.10${\tiny $\pm 0.24$} & \cellcolor{blue!30}$68.27${\tiny $\pm 0.48$} & \multicolumn{1}{l|}{\cellcolor{blue!30}$80.61${\tiny $\pm 0.19$}} & \cellcolor{blue!30}$90.80${\tiny $\pm 0.87$} & \cellcolor{blue!30}$90.32${\tiny $\pm 0.27$} & \cellcolor{blue!30}$82.48${\tiny $\pm 0.45$} & \cellcolor{blue!30}$89.94${\tiny $\pm 0.28$} \\ \bottomrule
\end{tabular}
}

\end{table*}

%% file: Table_SOTA_DCA1_XCAD.tex
\begin{table*}[t]
\caption{Comparison with state-of-the-arts on the \textit{DCA1} and \textit{XCAD} datasets. The best results are highlighted in \sethlcolor{blue!30}\hl{blue} and the second-best results are highlighted in \sethlcolor{red!30}\hl{red}. ``mean $\pm$ std'' is reported over three random seeds.}
\label{table:dca1_xcad}
\centering
\renewcommand\arraystretch{1.2}
\resizebox{\linewidth}{!}{
\begin{tabular}{lcccccccc}
\toprule
\multirow{2}{*}{Model} & \multicolumn{4}{c|}{\textit{DCA1}} & \multicolumn{4}{c}{\textit{XCAD}} \\ \cmidrule{2-9} 
& Sen. (\%) $\uparrow$ & F1 (\%) $\uparrow$ & IoU (\%) $\uparrow$ & \multicolumn{1}{l|}{MCC (\%) $\uparrow$} & Sen. (\%) $\uparrow$  & F1 (\%) $\uparrow$  & IoU (\%) $\uparrow$  & MCC (\%) $\uparrow$ \\ \midrule
UNet~\cite{ronneberger2015u}~{\tiny \color{gray}[MICCAI'15]} & $80.04${\tiny $\pm 0.36$} & $78.90${\tiny $\pm 0.21$} & $65.31${\tiny $\pm 0.30$} & \multicolumn{1}{l|}{$77.95${\tiny $\pm 0.23$}} & $80.64${\tiny $\pm 1.31$} & \cellcolor{red!30}$80.74${\tiny $\pm 0.08$} & \cellcolor{red!30}$67.98${\tiny $\pm 0.11$} & \cellcolor{red!30}$79.81${\tiny $\pm 0.12$} \\
UNet++~\cite{zhou2019unet++}~{\tiny \color{gray}[TMI'19]} & $80.10${\tiny $\pm 0.23$} & $78.45${\tiny $\pm 0.04$} & $64.69${\tiny $\pm 0.06$} & \multicolumn{1}{l|}{$77.49${\tiny $\pm 0.04$}} & $81.13${\tiny $\pm 0.62$} & $80.43${\tiny $\pm 0.25$} & $67.56${\tiny $\pm 0.34$} & $79.45${\tiny $\pm 0.26$} \\
AttnUNet~\cite{schlemper2019attention}~{\tiny \color{gray}[MedIA'19]} & $79.20${\tiny $\pm 0.30$} & $78.04${\tiny $\pm 0.01$} & $64.18${\tiny $\pm 0.01$} & \multicolumn{1}{l|}{$77.07${\tiny $\pm 0.01$}} & $80.08${\tiny $\pm 0.35$} & $79.98${\tiny $\pm 0.24$} & $66.94${\tiny $\pm 0.36$} & $79.01${\tiny $\pm 0.26$} \\
CE-Net~\cite{gu2019net}~{\tiny \color{gray}[TMI'19]} & $79.49${\tiny $\pm 0.73$} & $77.84${\tiny $\pm 0.12$} & $63.87${\tiny $\pm 0.16$} & \multicolumn{1}{l|}{$76.84${\tiny $\pm 0.13$}} & $79.96${\tiny $\pm 0.45$} & $79.95${\tiny $\pm 0.13$} & $66.80${\tiny $\pm 0.17$} & $78.95${\tiny $\pm 0.14$} \\
CAU-net~\cite{li2020cau}~{\tiny \color{gray}[ICONIP'20]} & $79.29${\tiny $\pm 0.18$} & $77.82${\tiny $\pm 0.12$} & $63.88${\tiny $\pm 0.15$} & \multicolumn{1}{l|}{$76.86${\tiny $\pm 0.12$}} & $80.22${\tiny $\pm 0.75$} & $79.31${\tiny $\pm 0.31$} & $66.06${\tiny $\pm 0.44$} & $78.31${\tiny $\pm 0.32$} \\
CS$^2$-Net~\cite{mou2021cs2}~{\tiny \color{gray}[MedIA'21]} & $78.46${\tiny $\pm 0.52$} & $77.87${\tiny $\pm 0.25$} & $63.94${\tiny $\pm 0.32$} & \multicolumn{1}{l|}{$76.92${\tiny $\pm 0.23$}} & $79.47${\tiny $\pm 0.72$} & $79.23${\tiny $\pm 0.30$} & $65.98${\tiny $\pm 0.39$} & $78.30${\tiny $\pm 0.32$} \\
FR-UNet~\cite{liu2022full}~{\tiny \color{gray}[JBHI'22]} & $78.96${\tiny $\pm 1.95$} & \cellcolor{red!30}$79.59${\tiny $\pm 0.30$} & \cellcolor{red!30}$66.22${\tiny $\pm 0.42$} & \multicolumn{1}{l|}{\cellcolor{blue!30}$79.47${\tiny $\pm 0.28$}} & \cellcolor{red!30}$81.65${\tiny $\pm 1.70$} & $79.79${\tiny $\pm 0.38$} & $66.66${\tiny $\pm 0.49$} & $78.87${\tiny $\pm 0.34$} \\
DE-DCGCN-EE~\cite{li2022dual}~{\tiny \color{gray}[TMI'22]} & $78.48${\tiny $\pm 0.19$} & $77.82${\tiny $\pm 0.09$} & $63.87${\tiny $\pm 0.11$} & \multicolumn{1}{l|}{$76.83${\tiny $\pm 0.08$}} & $79.89${\tiny $\pm 0.43$} & $79.12${\tiny $\pm 0.12$} & $65.76${\tiny $\pm 0.17$} & $78.06${\tiny $\pm 0.12$} \\
GT-DLA-dsHFF~\cite{li2023global}~{\tiny \color{gray}[TCYB'23]} & $75.62${\tiny $\pm 0.74$} & $77.17${\tiny $\pm 0.24$} & $62.97${\tiny $\pm 0.32$} & \multicolumn{1}{l|}{$76.12${\tiny $\pm 0.24$}} & $80.56${\tiny $\pm 0.52$} & $80.44${\tiny $\pm 0.35$} & $67.53${\tiny $\pm 0.44$} & $79.45${\tiny $\pm 0.36$} \\
TransUNet~\cite{chen2024transunet}~{\tiny \color{gray}[MedIA'24]} & \cellcolor{blue!30}$81.09${\tiny $\pm 0.46$} & $78.82${\tiny $\pm 0.12$} & $65.19${\tiny $\pm 0.18$} & \multicolumn{1}{l|}{$77.82${\tiny $\pm 0.12$}} & $80.86${\tiny $\pm 0.99$} & $80.24${\tiny $\pm 0.49$} & $67.28${\tiny $\pm 0.62$} & $79.26${\tiny $\pm 0.49$} \\
\midrule
SPIRONet~{\tiny \color{gray}\textbf{[Ours]}} & \cellcolor{red!30}$80.76${\tiny $\pm 0.94$} & \cellcolor{blue!30}$79.75${\tiny $\pm 0.48$} & \cellcolor{blue!30}$66.45${\tiny $\pm 0.66$} & \multicolumn{1}{l|}{\cellcolor{red!30}$78.75${\tiny $\pm 0.48$}} & \cellcolor{blue!30}$82.91${\tiny $\pm 0.63$} & \cellcolor{blue!30}$81.76${\tiny $\pm 0.51$} & \cellcolor{blue!30}$69.37${\tiny $\pm 0.69$} & \cellcolor{blue!30}$80.82${\tiny $\pm 0.53$} \\ 
\bottomrule
\end{tabular}
}

\end{table*}

%% file: Table_TTest.tex
\begin{table*}[t]
\centering
\caption{Paired image-level statistical analysis on the \textit{ARCADE} dataset. 
$\Delta$ denotes the mean per-image improvement of SPIRONet over each baseline. 
The 95\% confidence intervals are estimated by bootstrap resampling over testing images, and $p$-values are obtained using the two-sided Wilcoxon signed-rank test.}
\resizebox{\textwidth}{!}{
\begin{tabular}{lllllllll}
\toprule
\multirow{2}{*}{Model} 
& \multicolumn{2}{c}{Sen.} 
& \multicolumn{2}{c}{F1} 
& \multicolumn{2}{c}{IoU} 
& \multicolumn{2}{c}{MCC} \\
\cmidrule(lr){2-3} \cmidrule(lr){4-5} \cmidrule(lr){6-7} \cmidrule(lr){8-9}
& $\Delta$ ($95\%$ CI) & $p$-value
& $\Delta$ ($95\%$ CI) & $p$-value
& $\Delta$ ($95\%$ CI) & $p$-value
& $\Delta$ ($95\%$ CI) & $p$-value \\
\midrule
UNet~\cite{ronneberger2015u}~{\tiny \color{gray}[MICCAI'15]} & $7.22~[6.20, 8.27]$ & $p<0.001$
& $2.96~[2.20, 3.82]$ & $p<0.001$
& $3.35~[2.54, 4.24]$ & $p<0.001$
& $2.54~[1.84, 3.33]$ & $p<0.001$ \\

UNet++~\cite{zhou2019unet++}~{\tiny \color{gray}[TMI'19]} & $6.30~[5.31, 7.29]$ & $p<0.001$
& $2.89~[2.16, 3.62]$ & $p<0.001$
& $3.30~[2.51, 4.08]$ & $p<0.001$
& $2.55~[1.90, 3.20]$ & $p<0.001$ \\

AttnUNet~\cite{schlemper2019attention}~{\tiny \color{gray}[MedIA'19]} & $7.62~[6.59, 8.63]$ & $p<0.001$
& $3.40~[2.62, 4.21]$ & $p<0.001$
& $3.84~[3.02, 4.65]$ & $p<0.001$
& $2.93~[2.22, 3.68]$ & $p<0.001$ \\

CE-Net~\cite{gu2019net}~{\tiny \color{gray}[TMI'19]} & $5.79~[4.81, 6.80]$ & $p<0.001$
& $1.97~[1.23, 2.75]$ & $p<0.001$
& $2.22~[1.39, 3.06]$ & $p<0.001$
& $1.76~[0.92, 2.27]$ & $p<0.001$ \\

CAU-net~\cite{li2020cau}~{\tiny \color{gray}[ICONIP'20]} & $7.44~[6.45, 8.41]$ & $p<0.001$
& $3.70~[2.98, 4.44]$ & $p<0.001$
& $4.45~[3.66, 5.25]$ & $p<0.001$
& $3.43~[2.77, 4.09]$ & $p<0.001$ \\

CS$^2$-Net~\cite{mou2021cs2}~{\tiny \color{gray}[MedIA'21]} & $6.66~[5.72, 7.61]$ & $p<0.001$
& $3.66~[2.97, 4.36]$ & $p<0.001$
& $4.40~[3.64, 5.15]$ & $p<0.001$
& $3.47~[2.84, 4.11]$ & $p<0.001$ \\

FR-UNet~\cite{liu2022full}~{\tiny \color{gray}[JBHI'22]} & $2.42~[0.38,4.43]$ & $p<0.05$ 
& $7.49~[5.74,9.23]$ & $p<0.001$
& $9.17~[7.11,11.20]$ & $p<0.001$
& $7.73~[6.00,9.44]$ & $p<0.001$ \\

DE-DCGCN-EE~\cite{li2022dual}~{\tiny \color{gray}[TMI'22]} & $6.35~[5.57, 7.14]$ & $p<0.001$
& $4.04~[3.41, 4.69]$ & $p<0.001$
& $4.97~[4.28, 5.67]$ & $p<0.001$
& $4.03~[3.41, 4.66]$ & $p<0.001$ \\

GT-DLA-dsHFF~\cite{li2023global}~{\tiny \color{gray}[TCYB'23]} & $8.36~[7.43, 9.29]$ & $p<0.001$
& $3.69~[3.28, 4.75]$ & $p<0.001$
& $4.82~[4.05, 5.61]$ & $p<0.001$
& $3.70~[3.01, 4.43]$ & $p<0.001$ \\

TransUNet~\cite{chen2024transunet}~{\tiny \color{gray}[MedIA'24]} & $3.07~[2.29, 3.87]$ & $p<0.001$
& $1.95~[1.36, 2.54]$ & $p<0.001$
& $2.36~[1.56, 2.95]$ & $p<0.001$
& $1.94~[1.35, 2.53]$ & $p<0.001$ \\
\bottomrule
\end{tabular}
}
\label{tab:arcade_statistical_analysis}
\end{table*}

%% file: Figure_SOTA_Vis.tex
\begin{figure*}[t]
   \centering
   \centerline{\includegraphics{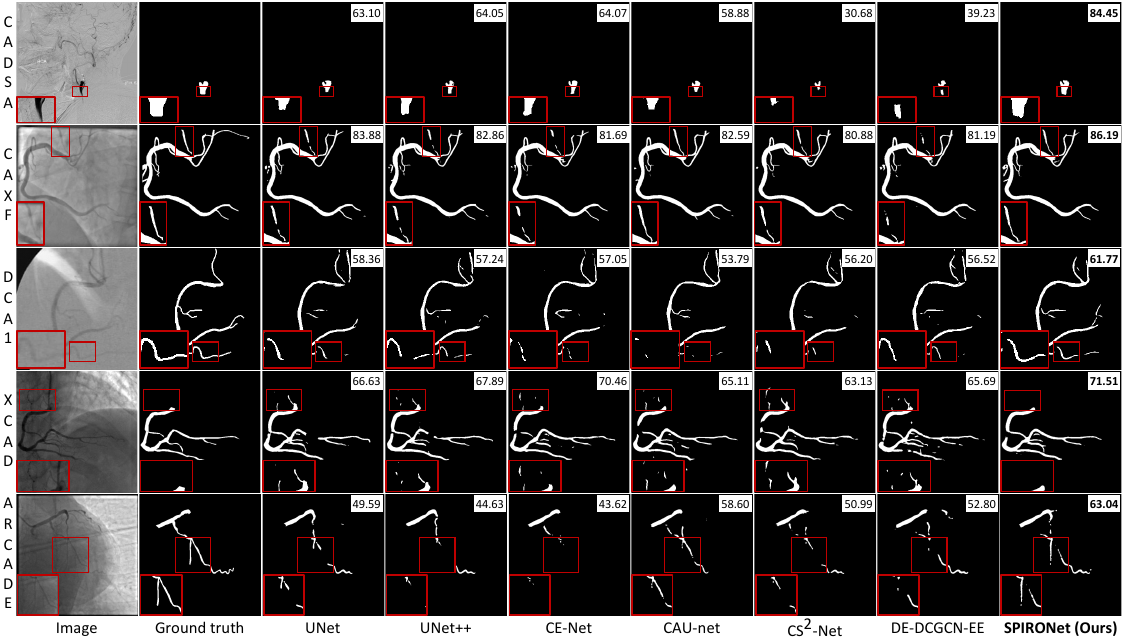}}
   \caption{Visualization of vessel segmentation results on the \textit{CADSA}, \textit{CAXF}~\cite{li2020cau}, \textit{DCA1}~\cite{cervantes2019automatic}, \textit{XCAD}~\cite{ma2021self}, and \textit{ARCADE}~\cite{popov2024dataset} datasets (from top to bottom). IoU (\%) for each image is shown in the upper right corner.}
   \label{fig:sota_compare}
\end{figure*}

%% file: Table_Ablation.tex
\begin{table*}[t]
\caption{Ablation results on the \textit{XCAD} and \textit{ARCADE} datasets. The best results are in \textbf{bold}. The second-best results are \underline{underlined}. All results are averaged over three random seeds.}
\label{table:ablation}
\centering
\renewcommand\arraystretch{1.2}
\resizebox{\linewidth}{!}{
\begin{tabular}{lcccc|cccc|cccc}
\toprule
\multirow{2}{*}{Variant} & \multicolumn{4}{c|}{Module} & \multicolumn{4}{c|}{\textit{XCAD}} & \multicolumn{4}{c}{\textit{ARCADE}} \\ \cmidrule{2-13} 
& SE & FE & CA & GCI & Sen. (\%) $\uparrow$ & F1 (\%) $\uparrow$ & IoU (\%) $\uparrow$ & MCC (\%) $\uparrow$ & Sen. (\%) $\uparrow$ & F1 (\%) $\uparrow$ & IoU (\%) $\uparrow$ & MCC (\%) $\uparrow$ \\ \midrule
I & {\color{mydeepgreen}\Checkmark} & {\color{red}\XSolidBrush} & {\color{red}\XSolidBrush} & {\color{red}\XSolidBrush} & $80.84$ & $80.06$ & $67.11$ & $79.11$ & $67.44$ & $72.93$ & $58.86$ & $73.04$  \\
II & {\color{red}\XSolidBrush} & {\color{mydeepgreen}\Checkmark} & {\color{red}\XSolidBrush} & {\color{red}\XSolidBrush} & $76.00$ & $77.14$ & $63.11$ & $76.19$ & $65.82$ & $71.89$ & $57.73$ & $72.18$ \\
III & {\color{mydeepgreen}\Checkmark} & {\color{mydeepgreen}\Checkmark} & {\color{red}\XSolidBrush} & {\color{red}\XSolidBrush} & $82.10$ & $81.09$ & $68.39$ & $80.15$ & $67.21$ & $73.69$ & $59.88$ & $74.00$ \\
IV & {\color{mydeepgreen}\Checkmark} & {\color{mydeepgreen}\Checkmark} & {\color{mydeepgreen}\Checkmark} & {\color{red}\XSolidBrush} & $82.00$ & $81.22$ & $68.57$ & $80.23$ & $\underline{71.46}$ & $\underline{76.22}$ & $\underline{62.78}$ & $\underline{76.19}$ \\
V & {\color{mydeepgreen}\Checkmark} & {\color{red}\XSolidBrush} & {\color{red}\XSolidBrush} & {\color{mydeepgreen}\Checkmark} & $82.23$ & $80.73$ & $67.98$ & $79.77$ & $69.45$ & $73.81$ & $59.82$ & $73.74$ \\ 
VI & {\color{red}\XSolidBrush} & {\color{mydeepgreen}\Checkmark} & {\color{red}\XSolidBrush} & {\color{mydeepgreen}\Checkmark} & $78.15$ & $77.63$ & $63.78$ & $76.64$ & $70.00$ & $72.57$ & $58.21$ & $72.41$ \\ 
VII & {\color{mydeepgreen}\Checkmark} & {\color{mydeepgreen}\Checkmark} & {\color{red}\XSolidBrush} & {\color{mydeepgreen}\Checkmark} & $\bm{83.07}$ & $\underline{81.36}$ & $\underline{68.76}$ & $\underline{80.40}$ & $67.65$ & $74.10$ & $60.28$ & $74.36$ \\
\midrule
SPIRONet & {\color{mydeepgreen}\Checkmark} & {\color{mydeepgreen}\Checkmark} & {\color{mydeepgreen}\Checkmark} & {\color{mydeepgreen}\Checkmark} & $\underline{82.91}$ & $\bm{81.76}$ & $\bm{69.37}$ & $\bm{80.82}$ & $\bm{75.65}$ & $\bm{77.15}$ & $\bm{63.91}$ & $\bm{76.88}$ \\
\bottomrule
\multicolumn{13}{l}{SE: Spatial Encoder; FE: Frequency Encoder; CA: Cross-Attention; {GCI: Graph-based Channel Interaction.}}
\end{tabular}
}

\end{table*}

%% file: Table_SOTA_ARCADE.tex
\begin{table}[htbp]
\caption{Comparison with state-of-the-arts on the \textit{ARCADE} dataset. The best results are highlighted in \sethlcolor{blue!30}\hl{blue} and the second-best results are highlighted in \sethlcolor{red!30}\hl{red}. ``mean $\pm$ std'' is reported over three random seeds.
}
\label{table:arcade}
\centering
\renewcommand\arraystretch{1.2}
\resizebox{\linewidth}{!}{
\begin{tabular}{lcccc}
\toprule
Model & Sen. (\%) $\uparrow$ & F1 (\%) $\uparrow$ & IoU (\%) $\uparrow$ & MCC (\%) $\uparrow$ \\ \midrule
UNet~\cite{ronneberger2015u}~{\tiny \color{gray}[MICCAI'15]} & $68.43${\tiny$\pm0.97$} & $74.19${\tiny$\pm0.33$} & $60.56${\tiny$\pm0.41$} & $74.34${\tiny$\pm0.20$} \\
UNet++~\cite{zhou2019unet++}~{\tiny \color{gray}[TMI'19]} & $69.35${\tiny$\pm0.18$} & $74.26${\tiny$\pm0.20$} & $60.61${\tiny$\pm0.24$} & $74.33${\tiny$\pm0.20$} \\
AttnUNet~\cite{schlemper2019attention}~{\tiny \color{gray}[MedIA'19]} & $68.03${\tiny$\pm0.21$} & $73.75${\tiny$\pm0.11$} & $60.07${\tiny$\pm0.08$} & $73.95${\tiny$\pm0.11$} \\
CE-Net~\cite{gu2019net}~{\tiny \color{gray}[TMI'19]} & $69.86${\tiny$\pm0.41$} & $75.18${\tiny$\pm0.28$} & \cellcolor{red!30}$61.69${\tiny$\pm0.40$} & \cellcolor{red!30}$75.12${\tiny$\pm0.55$} \\
CAU-net~\cite{li2020cau}~{\tiny \color{gray}[ICONIP'20]} & $68.21${\tiny$\pm0.78$} & $73.45${\tiny$\pm0.39$} & $59.46${\tiny$\pm0.49$} & $73.45${\tiny$\pm0.34$} \\
CS$^2$-Net~\cite{mou2021cs2}~{\tiny \color{gray}[MedIA'21]} & $68.99${\tiny$\pm0.63$} & $73.49${\tiny$\pm0.66$} & $59.51${\tiny$\pm0.80$} & $73.41${\tiny$\pm0.65$} \\
FR-UNet~\cite{liu2022full}~{\tiny \color{gray}[JBHI'22]} & \cellcolor{red!30}$73.23${\tiny$\pm0.11$} & $69.66${\tiny$\pm0.03$} & $54.74${\tiny$\pm0.04$} & $69.15${\tiny$\pm0.03$} \\
DE-DCGCN-EE~\cite{li2022dual}~{\tiny \color{gray}[TMI'22]} & $69.30${\tiny$\pm1.12$} & $73.11${\tiny$\pm0.48$} & $58.94${\tiny$\pm0.58$} & $72.85${\tiny$\pm0.44$} \\
GT-DLA-dsHFF~\cite{li2023global}~{\tiny \color{gray}[TCYB'23]} & $67.29${\tiny$\pm1.82$} & $73.46${\tiny$\pm0.33$} & $59.09${\tiny$\pm0.90$} & $73.18${\tiny$\pm0.57$} \\
TransUNet~\cite{chen2024transunet}~{\tiny \color{gray}[MedIA'24]} & $72.58${\tiny$\pm1.37$} & \cellcolor{red!30}$75.20${\tiny$\pm0.73$} & $61.55${\tiny$\pm0.98$} & $74.94${\tiny$\pm0.83$} \\
\midrule
SPIRONet~{\tiny \color{gray}\textbf{[Ours]}} & \cellcolor{blue!30}$75.65${\tiny$\pm1.58$} & \cellcolor{blue!30}$77.15${\tiny$\pm0.15$} & \cellcolor{blue!30}$63.91${\tiny$\pm0.24$} & \cellcolor{blue!30}$76.88${\tiny$\pm0.19$} \\
\bottomrule
\end{tabular}
}
\end{table}

%% file: Table_TTest_All_Datasets.tex
\begin{table}[htbp]
\caption{Paired image-level statistical analysis between SPIRONet and the strongest competing method on each dataset. $\Delta$ denotes the mean per-image IoU improvement of SPIRONet over the strongest competing method. The 95\% confidence intervals are estimated by bootstrap resampling over testing images, and $p$-values are obtained using the two-sided Wilcoxon signed-rank test.}
\label{table:wilcoxon_all_datasets}
\centering
\renewcommand\arraystretch{1.2}
\resizebox{\linewidth}{!}{
\begin{tabular}{llll}
\toprule
Dataset & Method & $\Delta$ IoU (95\% CI) & $p$-value \\
\midrule
\textit{CADSA} & CAU-net~\cite{li2020cau}~{\tiny \color{gray}[ICONIP'20]} 
& $0.87$ [$-0.58$, $2.51$] & $p=0.44$ \\
\textit{CAXF} & UNet++~\cite{zhou2019unet++}~{\tiny \color{gray}[TMI'19]} 
& $0.52$ [$0.24$, $0.87$] & $p<0.001$ \\
\textit{DCA1} & FR-UNet~\cite{liu2022full}~{\tiny \color{gray}[JBHI'22]} 
& $0.23$ [$-0.05$, $2.44$] & $p=0.05$ \\
\textit{XCAD} & UNet~\cite{ronneberger2015u}~{\tiny \color{gray}[MICCAI'15]} 
& $1.39$ [$0.30$, $2.20$] & $p<0.05$ \\
\textit{ARCADE} & CE-Net~\cite{gu2019net}~{\tiny \color{gray}[TMI'19]} 
& $2.22$ [$1.39$, $3.06$] & $p<0.001$ \\
\bottomrule
\end{tabular}
}
\end{table}

%% file: Table_Param.tex
\begin{table}[htbp]
\caption{Model parameters, floating-point operations (FLOPs), and inference rates. The input size of models is set to $512\times512$. $^*$ means the experiments are conducted on the \textit{test} set of the \textit{XCAD} dataset and repeated five times.
}
\label{table:complexity}
\centering
\renewcommand\arraystretch{1.2}
\resizebox{\linewidth}{!}{
\begin{tabular}{llll}
\toprule
Model & \#params (M) & FLOPs (G) & Rate$^*$ (FPS) \\ \midrule
UNet~\cite{ronneberger2015u}~{\tiny \color{gray}[MICCAI'15]} & $17.26$ & $160.44$ & $61.52${\tiny$\pm0.91$} \\
UNet++~\cite{zhou2019unet++}~{\tiny \color{gray}[TMI'19]} & $9.16$ & $139.46$ & $45.03${\tiny$\pm0.08$} \\
AttnUNet~\cite{schlemper2019attention}~{\tiny \color{gray}[MedIA'19]} & $57.16$ & $541.04$ & $29.00${\tiny$\pm0.26$} \\
CE-Net~\cite{gu2019net}~{\tiny \color{gray}[TMI'19]} & $29.00$ & $35.60$ & $74.11${\tiny$\pm0.38$} \\
CAU-net~\cite{li2020cau}~{\tiny \color{gray}[ICONIP'20]} & $1.95$ & $13.91$ & $102.98${\tiny$\pm5.37$} \\
CS$^2$-Net~\cite{mou2021cs2}~{\tiny \color{gray}[MedIA'21]} & $8.40$ & $55.85$ & $82.02${\tiny$\pm0.43$} \\
FR-UNet~\cite{liu2022full}~{\tiny \color{gray}[JBHI'22]} & $5.72$ & $235.60$ & $26.75${\tiny$\pm0.05$} \\
DE-DCGCN-EE~\cite{li2022dual}~{\tiny \color{gray}[TMI'22]} & $14.11$ & $294.46$ & $18.07${\tiny$\pm0.02$} \\
GT-DLA-dsHFF~\cite{li2023global}~{\tiny \color{gray}[TCYB'23]} & $26.10$ & $474.60$ & $11.80${\tiny$\pm0.20$} \\ 
TransUNet~\cite{chen2024transunet}~{\tiny \color{gray}[MedIA'24]} & $93.23$ & $129.45$ & $24.82${\tiny$\pm0.13$} \\
\midrule
SPIRONet~{\tiny \color{gray}\textbf{[Ours]}} & $16.98$ & $200.90$ & $20.57${\tiny$\pm0.09$}\\
\bottomrule
\end{tabular}
}
\end{table}

%% file: Table_FrequencyFilter.tex
\begin{table}[htbp]
\caption{
Comparison with other frequency operators on the \textit{XCAD} dataset. The best results are in \textbf{bold}. The second-best results are \underline{underlined}. All results are averaged over three random seeds.
}
\label{table:frequencyfilter}
\centering
\renewcommand\arraystretch{1.2}
\resizebox{\linewidth}{!}{
\begin{tabular}{lllll}
\toprule
Model & Sen. (\%) $\uparrow$  & F1 (\%) $\uparrow$  & IoU (\%) $\uparrow$  & MCC (\%) $\uparrow$ \\ \midrule
V {\tiny\color{gray}\textbf{[Ours]}} & $82.23$ & $80.73$ & $67.98$ & $79.77$ \\
\ +AFNO~\cite{guibas2021adaptive} {\tiny\color{gray}[ICLR'22]} & $82.22$ {\color{red}$\downarrow 0.01$} & $80.60$ {\color{red}$\downarrow 0.13$} & $67.80$ {\color{red}$\downarrow 0.18$} & $79.67$ {\color{red}$\downarrow 0.10$} \\
\ +GFN~\cite{rao2023gfnet} {\tiny\color{gray}[TPAMI'23]} & $82.14$ {\color{red}$\downarrow 0.09$} & $80.62$ {\color{red}$\downarrow 0.11$} & $67.82$ {\color{red}$\downarrow 0.16$} & $79.67$ {\color{red}$\downarrow 0.10$} \\
\ +FE {\tiny\color{gray}\textbf{[Ours]}} & $\underline{83.07}$ {\color{mydeepgreen}$\uparrow 0.84$} & $\underline{81.36}$ {\color{mydeepgreen}$\uparrow 0.63$} & $\underline{68.76}$ {\color{mydeepgreen}$\uparrow 0.78$} & $\underline{80.40}$ {\color{mydeepgreen}$\uparrow 0.63$} \\
\ +CA+AFNO~\cite{guibas2021adaptive} {\tiny\color{gray}[ICLR'22]} & $83.05$ {\color{mydeepgreen}$\uparrow 0.82$} & $81.23$ {\color{mydeepgreen}$\uparrow 0.50$} & $68.62$ {\color{mydeepgreen}$\uparrow 0.64$} & $80.24$ {\color{mydeepgreen}$\uparrow 0.47$} \\
\ +CA+GFN~\cite{rao2023gfnet} {\tiny\color{gray}[TPAMI'23]} & $\bm{83.12}$ {\color{mydeepgreen}$\uparrow\bm{0.89}$} & $81.24$ {\color{mydeepgreen}$\uparrow 0.51$} & $68.61$ {\color{mydeepgreen}$\uparrow 0.63$} & $80.26$ {\color{mydeepgreen}$\uparrow 0.49$} \\
\midrule
\rowcolor{gray!20}\ +CA+FE {\tiny\color{gray}\textbf{[Ours]}} & $82.91$ {\color{mydeepgreen}$\uparrow 0.68$} & $\bm{81.76}$ {\color{mydeepgreen}$\uparrow\bm{1.03}$} & $\bm{69.37}$ {\color{mydeepgreen}$\uparrow\bm{1.39}$} & $\bm{80.82}$ {\color{mydeepgreen}$\uparrow\bm{1.05}$} \\ \bottomrule
\multicolumn{5}{l}{FE: Frequency Encoder; CA: Cross-Attention.}
\end{tabular}
}
\end{table}

%% file: sec6.tex
In this section, a detailed analysis of the proposed SPIRONet is presented. Default configurations of SPIRONet are highlighted in \sethlcolor{light-gray}\hl{gray}. All experiments are conducted on the \textit{XCAD} dataset unless stated otherwise.

\subsection{Frequency Encoder}
\input{Table_FE_Pha_Amp}
\textbf{Comparison with Other Frequency Operators.} Given the critical role of the frequency encoder, the design of its frequency operation is further analyzed against approaches from previous works~\cite{guibas2021adaptive},~\cite{rao2023gfnet}. Quantitative results are presented in Table~\ref{table:frequencyfilter}. The baseline refers to the model without frequency encoder blocks (Variant V in Table~\ref{table:ablation}). It should be noted that in models incorporating frequency features but lacking the cross-attention module, element-wise addition is performed to fuse spatial and frequency features. Several observations can be drawn from Table~\ref{table:frequencyfilter}: \textbf{\textit{(i)}} When equipped with the cross-attention module, the models incorporating frequency features show consistent performance improvements. \textbf{\textit{(ii)}} The proposed frequency operation design clearly outperforms those in AFNO~\cite{guibas2021adaptive} and GFN~\cite{rao2023gfnet}, which utilize deterministic functions or masks to filter frequency information. In contrast, SPIRONet efficiently learns input-adaptive frequency features, benefiting from the semantic-adaptive design of the proposed frequency encoder.

\textbf{Roles of Amplitude and Phase Components.} To further investigate the roles of amplitude and phase components in the frequency encoder, we conduct an ablation study on the \textit{XCAD} dataset, as shown in Table~\ref{table:amp_pha}. In the amplitude-only and phase-only variants, the learnable convolutional path modulates only the amplitude and phase spectra, respectively, while discarding the other component during reconstruction. The results show that both amplitude and phase components contribute to vessel segmentation. Compared with the baseline, using only the phase component improves IoU by $0.75\%$, indicating that phase information is useful for preserving vessel-related spatial structures. Using only the amplitude component achieves a larger IoU improvement of $0.95\%$, suggesting that amplitude information provides discriminative spectral cues related to vessel contrast and texture. When both components are used together, the model achieves the best overall performance. These results demonstrate that amplitude and phase components provide complementary information, and their joint modeling enables the frequency encoder to learn more effective vessel-related spectral representations under low-SNR conditions.

\subsection{Graph-based Channel Interaction}
\input{Table_ChannelRefine}
\textbf{Comparison with Channel and Spatial Attention Modules.} To verify the effectiveness of the graph-based channel interaction (GCI) module, it is compared with several well-established channel refinement modules and spatial attention modules. Table~\ref{table:channel} summarizes the results. GCI demonstrates superior performance, outperforming not only other channel refinement modules but also spatial attention modules, underscoring the importance of directly modeling interactions among channel features. An interesting finding is that adding conventional channel refinement modules to our baseline (Variant IV) actually degraded performance. This phenomenon strongly suggests that simply re-weighting channels can be suboptimal for vessel segmentation, potentially introducing extra interference. Instead of merely allocating weights, GCI models the unstructured relationships between channels and facilitates direct information exchange via GNNs. This allows it to effectively filter irrelevant noise and enhance vessel-specific responses, leading to considerable performance gains, \textit{i.e.,} $0.87\%\sim1.11\%$ IoU improvements over other channel refinement modules.\input{Figure_GradCAM}\input{Figure_GNN_Analysis}\input{Table_Adjacent_Matrix_Construct}Additionally, Grad-CAM~\cite{selvaraju2017grad} is employed to compare discriminative regions with and without channel refinement modules. As depicted in Fig.~\ref{fig:gradcam}, regions highlighted in red boxes exhibit serious interference, including vessel-like interventional instruments and motion artifacts. The comparison methods tend to assign weights to these irrelevant regions (highlighted in red boxes), thus producing more false-positive predictions. By leveraging the graph-based channel interaction module, SPIRONet demonstrates an enhanced ability to filter out irrelevant responses compared with other methods.

\textbf{Comparison with Different Adjacency Matrix Construction Strategies.} To further validate the contribution of channel-node interaction in GCI, we compare different adjacency matrix construction strategies. Specifically, the identity matrix only preserves self-connections for each channel node and disables inter-channel interaction. The uniform fully-connected matrix connects all channel nodes with equal weights. The learnable matrix uses a trainable but input-independent adjacency matrix shared by all images, which tests whether a static graph topology can model channel relationships effectively. In contrast, the cosine similarity strategy used in GCI dynamically constructs the adjacency matrix according to the input features, enabling sample-specific channel-node interaction.\input{Table_PPM}As shown in Table~\ref{table:adjacent_matrix_construct}, the cosine similarity strategy achieves the best performance across all metrics. Compared with the identity matrix, it improves IoU by 0.67\%, indicating that enabling adaptive inter-channel interaction is more effective than preserving only independent channel responses. The uniform fully-connected matrix achieves better performance than the identity matrix, suggesting that inter-channel connections are beneficial. However, it still underperforms the cosine similarity strategy, showing that treating all channel pairs equally is suboptimal. Moreover, the learnable matrix obtains 68.52\% IoU, which is lower than the proposed strategy, indicating that a static adjacency matrix shared across all inputs is insufficient to capture image-specific channel relationships. These results demonstrate that the performance gain of GCI mainly comes from input-adaptive channel-node interaction rather than simply introducing additional parameters or uniformly mixing channel features.

\textbf{Interpreting Graph-based Channel Interaction.} To further interpret the mechanism of the proposed GCI, we visualize the learned adjacency matrix and the feature responses before and after GCI, as shown in Fig.~\ref{fig:gnn_analysis}. The adjacency matrix exhibits clear non-uniform connection patterns among channel nodes, indicating that different channels are not treated independently. Instead, GCI constructs a channel correlation graph and enables information exchange among correlated channel responses. This supports the motivation of modeling channel-wise dependencies through graph reasoning rather than simply recalibrating individual channels. The Grad-CAM visualization further illustrates the effect of GCI on feature refinement. Before GCI, the activation map is relatively weak and scattered, and some vessel regions are not sufficiently highlighted. After GCI, the discriminative responses become more concentrated along the vessel structures, especially in thin and low-contrast regions. These observations indicate that GCI enhances vessel-related channel responses while weakening task-irrelevant activations caused by noise or vessel-like interference.

\subsection{Analysis of Pyramid Pooling Modules}
Since PPM is introduced into the cross-attention module to reduce the computational cost of matrix multiplication, it is necessary to examine whether such spatial sampling affects small and slender vessel segmentation. Therefore, we further analyze different PPM configurations on the \textit{XCAD} dataset. In addition to IoU, clDice~\cite{shit2021cldice} is reported to evaluate the preservation of vessel centerlines and tubular structures, which is related to the segmentation quality of thin vessels. Moreover, we report the attention complexity rather than the whole-model FLOPs, because the total FLOPs are dominated by convolutional operations and are therefore insensitive to changes in the number of key/value tokens within the cross-attention module.

As shown in Table~\ref{table:ppm}, full attention without PPM is computationally prohibitive under the 512$\times$512 input setting. Specifically, without PPM, all spatial positions are used as key/value tokens, resulting in an attention complexity of $\mathcal{O}(M_i^2d_0)$, where $M_i=H_iW_i$ denotes the number of query tokens at the $i$-th scale and $d_0$ denotes the embedding dimension. This leads to out-of-memory errors on both an NVIDIA RTX $4090$ ($24$GB) and an NVIDIA RTX A$6000$ ($48$GB), indicating the necessity of reducing the number of key/value tokens in the cross-attention module. By contrast, PPM substantially reduces the number of key/value tokens from $M_i$ to $N$, and the attention complexity is reduced to $\mathcal{O}(M_iNd_0)$, where $N \ll M_i$. When the pooling scales are set to $\{1,3,6\}$, only 46 tokens are sampled and the model achieves 69.04\% IoU and 85.44\% clDice. Increasing the scales to $\{1,3,6,8\}$, which is the default setting in SPIRONet, increases the sampled tokens to 110 and improves IoU to 69.37\%, while maintaining the same clDice of 85.44\%. Further increasing the number of tokens to 254 with scales $\{1,3,6,8,12\}$ does not bring additional benefits. These results indicate that the default PPM setting $\{1,3,6,8\}$ provides a favorable trade-off between computational feasibility and vessel detail preservation. Since PPM is applied only to the key and value branches while the query branch retains dense spatial resolution, the cross-attention output is still computed for each spatial location. This design reduces attention complexity while preserving fine vessel responses, as reflected by the stable clDice performance.

%% file: Table_FE_Pha_Amp.tex
\begin{table}[htbp]
\caption{
Impact of the amplitude and phase components on the \textit{XCAD} dataset. The best results are in \textbf{bold}. The second-best results are \underline{underlined}. All results are averaged over three random seeds.
}
\label{table:amp_pha}
\centering
\renewcommand\arraystretch{1.2}
\resizebox{\linewidth}{!}{
\begin{tabular}{llllll}
\toprule
\multicolumn{2}{c}{Components} & \multirow{2}{*}{Sen. (\%) $\uparrow$} & \multirow{2}{*}{F1 (\%) $\uparrow$} & \multirow{2}{*}{IoU (\%) $\uparrow$} & \multirow{2}{*}{MCC (\%) $\uparrow$} \\ \cmidrule{1-2}
Amplitude & Phase & & & & \\ \midrule
{\color{red}\XSolidBrush} & {\color{red}\XSolidBrush} & $82.23$ & $80.73$ & $67.98$ & $79.77$ \\
{\color{red}\XSolidBrush} & {\color{mydeepgreen}\Checkmark} & $\bm{83.13}$ {\color{mydeepgreen}$\bm{\uparrow 0.90}$} & $81.31$ {\color{mydeepgreen}$\uparrow 0.58$} & $68.73$ {\color{mydeepgreen}$\uparrow 0.75$} & $80.33$ {\color{mydeepgreen}$\uparrow 0.56$} \\
{\color{mydeepgreen}\Checkmark} & {\color{red}\XSolidBrush} & $\underline{82.96}$ {\color{mydeepgreen}$\uparrow 0.73$} & $\underline{81.48}$ {\color{mydeepgreen}$\uparrow 0.75$} & $\underline{68.93}$ {\color{mydeepgreen}$\uparrow 0.95$} & $\underline{80.53}$ {\color{mydeepgreen}$\uparrow 0.76$} \\
\rowcolor{gray!20} {\color{mydeepgreen}\Checkmark} & {\color{mydeepgreen}\Checkmark} & $82.91$ {\color{mydeepgreen}$\uparrow 0.68$} & $\bm{81.76}$ {\color{mydeepgreen}$\bm{\uparrow 1.03}$} & $\bm{69.37}$ {\color{mydeepgreen}$\bm{\uparrow 1.39}$} & $\bm{80.82}$ {\color{mydeepgreen}$\bm{\uparrow 1.05}$} \\ \bottomrule
\end{tabular}
}
\end{table}

%% file: Table_ChannelRefine.tex
\begin{table}[htbp]
\caption{Comparison with other channel refinement modules and spatial attention modules on the \textit{XCAD} dataset. The best results are in \textbf{bold}. The second-best results are \underline{underlined}. All results are averaged over three random seeds.
}
\label{table:channel}
\centering
\renewcommand\arraystretch{1.2}
\resizebox{\linewidth}{!}{
\begin{tabular}{lllll}
\toprule
Model & Sen. (\%) $\uparrow$  & F1 (\%) $\uparrow$  & IoU (\%) $\uparrow$  & MCC (\%) $\uparrow$ \\ \midrule
 IV {\tiny\color{gray}\textbf{[Ours]}} & $82.00$ & $81.22$ & $68.57$ & $80.23$ \\
\ +Non-Local$^\dagger$~\cite{wang2018non} {\tiny\color{gray}[CVPR'18]} & $82.13$ {\color{mydeepgreen}$\uparrow0.13$} & $81.29$ {\color{mydeepgreen}$\uparrow0.07$} & $68.67$ {\color{mydeepgreen}$\uparrow0.10$} & \underline{$80.31$} {\color{mydeepgreen}$\uparrow0.08$} \\ 
\ +SE~\cite{hu2019squeeze} {\tiny\color{gray}[TPAMI'19]} & $81.61$ {\color{red}$\downarrow0.39$} & $80.99$ {\color{red}$\downarrow0.23$} & $68.26$ {\color{red}$\downarrow0.31$} & $80.02$ {\color{red}$\downarrow0.21$} \\
\ +SK~\cite{li2019selective} {\tiny\color{gray}[CVPR'19]} & $81.77$ {\color{red}$\downarrow0.23$} & $81.06$ {\color{red}$\downarrow0.16$} & $68.35$ {\color{red}$\downarrow0.22$} & $80.07$ {\color{red}$\downarrow0.16$} \\
\ +ECA~\cite{wang2020eca} {\tiny\color{gray}[CVPR'20]} & $81.66$ {\color{red}$\downarrow0.34$} & $81.06$ {\color{red}$\downarrow0.16$} & $68.34$ {\color{red}$\downarrow0.23$} & $80.06$ {\color{red}$\downarrow0.17$} \\
\ +MHA$^\dagger$~\cite{dosovitskiy2020image} {\tiny\color{gray}[ICLR'21]} & \underline{$82.20$} {\color{mydeepgreen}$\uparrow0.20$} & \underline{$81.31$} {\color{mydeepgreen}$\uparrow0.09$} & \underline{$68.70$} {\color{mydeepgreen}$\uparrow0.13$} & \underline{$80.31$} {\color{mydeepgreen}$\uparrow0.08$} \\
\ +FCA~\cite{qin2021fcanet} {\tiny\color{gray}[ICCV'21]} & $81.58$ {\color{red}$\downarrow0.42$} & $81.17$ {\color{red}$\downarrow0.05$} & $68.50$ {\color{red}$\downarrow0.07$} & $80.19$ {\color{red}$\downarrow0.04$} \\
\midrule
\rowcolor{gray!20}\ +GCI {\tiny\color{gray}\textbf{[Ours]}} & $\bm{82.91}$ {\color{mydeepgreen}$\uparrow\bm{0.91}$} & $\bm{81.76}$ {\color{mydeepgreen}$\uparrow\bm{0.54}$} & $\bm{69.37}$ {\color{mydeepgreen}$\uparrow\bm{0.80}$} & $\bm{80.82}$ {\color{mydeepgreen}$\uparrow\bm{0.59}$} \\ \bottomrule
\multicolumn{5}{l}{GCI: Graph-based Channel Interaction; $\dagger$: Spatial Attention Module.}
\end{tabular}
}
\end{table}

%% file: Figure_GradCAM.tex
\begin{figure}[htbp]
    \centerline{\includegraphics{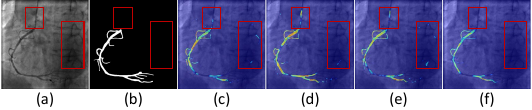}}
    \caption{Visualization of class activation maps~\cite{selvaraju2017grad} produced by different channel refinement modules. (a) Image. (b) Ground truth. (c) IV. (d) IV+SK~\cite{li2019selective}. (e) IV+FCA~\cite{qin2021fcanet}. (f) IV+GCI \textbf{(Ours)}.}
    \label{fig:gradcam}
\end{figure}

%% file: Figure_GNN_Analysis.tex
\begin{figure}[htbp]
    \centerline{\includegraphics{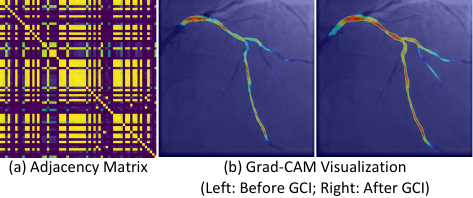}}
    \caption{Visualization of the graph-based channel interaction module. (a) Learned adjacency matrix among channel nodes. (b) Grad-CAM visualization before and after GCI.}
    \label{fig:gnn_analysis}
\end{figure}

%% file: Table_Adjacent_Matrix_Construct.tex
\begin{table}[htbp]
\caption{Ablation study of different adjacency matrix construction strategies in GCI on the \textit{XCAD} dataset. The best results are in \textbf{bold}. The second-best results are \underline{underlined}. All results are averaged over three random seeds.}
\label{table:adjacent_matrix_construct}
\centering
\renewcommand\arraystretch{1.2}
\resizebox{\linewidth}{!}{
\begin{tabular}{lllll}
\toprule
Strategy & Sen. (\%) $\uparrow$ & F1 (\%) $\uparrow$ & IoU (\%) $\uparrow$ & MCC (\%) $\uparrow$ \\
\midrule
Identity Matrix & $\underline{82.28}$ & $81.31$ & $68.70$ & $80.32$ \\
Uniform Fully-connected Matrix & $82.25$ & $\underline{81.44}$ & $\underline{68.89}$ & $\underline{80.48}$ \\
Learnable Matrix & $82.15$ & $81.16$ & $68.52$ & $80.19$ \\
\rowcolor{gray!20}Cosine Similarity & $\bm{82.91}$ & $\bm{81.76}$ & $\bm{69.37}$ & $\bm{80.82}$ \\
\bottomrule
\end{tabular}
}
\end{table}

%% file: Table_PPM.tex
\begin{table}[htbp]
\caption{
Effect of PPM configurations on the \textit{XCAD} dataset. The best results are in \textbf{bold}. The second-best results are \underline{underlined}. All results are averaged over three random seeds. $M_i=H_iW_i$ denotes the number of query tokens at the $i$-th scale, and $N$ denotes the number of key/value tokens sampled by PPM. OOM: Out-of-memory. Full attention runs out of memory on both NVIDIA RTX $4090$ ($24$GB) and NVIDIA RTX A$6000$ ($48$GB).
}
\label{table:ppm}
\centering
\renewcommand\arraystretch{1.2}
\resizebox{\linewidth}{!}{
\begin{tabular}{lllll}
\toprule
PPM Scales & Tokens $N$ & Attn. Complexity & IoU (\%) $\uparrow$ & clDice (\%) $\uparrow$ \\
\midrule
Full Attention & $H_iW_i$ & $\mathcal{O}\left(M_i^2d_0\right)$ & OOM & OOM \\
$\{1,3,6\}$ & $46$ & $\mathcal{O}\left(46M_id_0\right)$ & $69.04$ & $\bm{85.44}$ \\
\rowcolor{gray!20}$\{1,3,6,8\}$ & $110$ & $\mathcal{O}(110M_id_0)$ & $\bm{69.37}$ & $\bm{85.44}$ \\
$\{1,3,6,8,12\}$ & $254$ & $\mathcal{O}(254M_id_0)$ & $\underline{69.15}$ & $85.36$ \\
\bottomrule
\end{tabular}
}
\end{table}

%% file: sec7.tex
This paper proposes a \underline{\textbf{SP}}atial-frequency learning and graph-based channel \underline{\textbf{I}}nte\underline{\textbf{R}}acti\underline{\textbf{O}}n \underline{\textbf{Net}}work (\textbf{SPIRONet}) to tackle challenges in vessel segmentation. The dual encoders in SPIRONet effectively extract local spatial and global frequency vessel features. The complementary features are then fused using cross-attention fusion modules. Furthermore, the proposed graph-based channel interaction module is able to filter out task-irrelevant responses in multi-channel feature maps. Experimental results on five benchmarks demonstrate the effectiveness of the proposed model. In future work, knowledge distillation will be explored to further enhance the inference efficiency of SPIRONet. Additionally, SPIRONet will be validated in real-world clinical scenarios.

%% file: cas-dc-template.bbl
\begin{thebibliography}{77}
\expandafter\ifx\csname natexlab\endcsname\relax\def\natexlab#1{#1}\fi
\providecommand{\url}[1]{\texttt{#1}}
\providecommand{\href}[2]{#2}
\providecommand{\path}[1]{#1}
\providecommand{\DOIprefix}{doi:}
\providecommand{\ArXivprefix}{arXiv:}
\providecommand{\URLprefix}{URL: }
\providecommand{\Pubmedprefix}{pmid:}
\providecommand{\doi}[1]{\href{http://dx.doi.org/#1}{\path{#1}}}
\providecommand{\Pubmed}[1]{\href{pmid:#1}{\path{#1}}}
\providecommand{\bibinfo}[2]{#2}
\ifx\xfnm\relax \def\xfnm[#1]{\unskip,\space#1}\fi
\bibitem[{Sacco(2016)}]{sacco2016heart}
\bibinfo{author}{R.~L.~o. Sacco},
\newblock \bibinfo{title}{The heart of 25 by 25: Achieving the goal of reducing global and regional premature deaths from cardiovascular diseases and stroke: A modeling study from the american heart association and world heart federation},
\newblock \bibinfo{journal}{Circulation} \bibinfo{volume}{133} (\bibinfo{year}{2016}) \bibinfo{pages}{e674--e690}.
\bibitem[{Roth et~al.(2020)Roth, Mensah, and Fuster}]{roth2020global}
\bibinfo{author}{G.~A. Roth}, \bibinfo{author}{G.~A. Mensah}, \bibinfo{author}{V.~Fuster},
\newblock \bibinfo{title}{The global burden of cardiovascular diseases and risks: A compass for global action},
\newblock \bibinfo{journal}{J. Am. Coll. Cardiol.} \bibinfo{volume}{76} (\bibinfo{year}{2020}) \bibinfo{pages}{2980--2981}.
\bibitem[{Wan et~al.(2022)}]{wan2022symptomatic}
\bibinfo{author}{M.~Wan}, et~al.,
\newblock \bibinfo{title}{Symptomatic and asymptomatic chronic carotid artery occlusion on high-resolution {MR} vessel wall imaging},
\newblock \bibinfo{journal}{Am. J. Neuroradiol.} \bibinfo{volume}{43} (\bibinfo{year}{2022}) \bibinfo{pages}{110--116}.
\bibitem[{Langer and Argenziano(2016)}]{langer2016minimally}
\bibinfo{author}{N.~B. Langer}, \bibinfo{author}{M.~Argenziano},
\newblock \bibinfo{title}{Minimally invasive cardiovascular surgery: Incisions and approaches},
\newblock \bibinfo{journal}{Methodist Debakey Cardiovasc. J.} \bibinfo{volume}{12} (\bibinfo{year}{2016}) \bibinfo{pages}{4}.
\bibitem[{Cagnazzo et~al.(2020)}]{cagnazzo2020endovascular}
\bibinfo{author}{F.~Cagnazzo}, et~al.,
\newblock \bibinfo{title}{Endovascular recanalization of chronically occluded internal carotid artery},
\newblock \bibinfo{journal}{J. Neurointerv. Surg.} \bibinfo{volume}{12} (\bibinfo{year}{2020}) \bibinfo{pages}{946--951}.
\bibitem[{Li et~al.(2023)Li, Zhou, Xie, Liu, Feng, and Hou}]{hao2023casog}
\bibinfo{author}{H.~Li}, \bibinfo{author}{X.-H. Zhou}, \bibinfo{author}{X.-L. Xie}, \bibinfo{author}{S.-Q. Liu}, \bibinfo{author}{Z.-Q. Feng}, \bibinfo{author}{Z.-G. Hou},
\newblock \bibinfo{title}{{CASOG}: Conservative actor–critic with smooth gradient for skill learning in robot-assisted intervention},
\newblock \bibinfo{journal}{IEEE Trans. Ind. Electron.}  (\bibinfo{year}{2023}). \bibinfo{note}{{D}OI:{\color{blue} \href{http://dx.doi.org/10.1109/TIE.2023.3310021}{10.1109/TIE.2023.3310021}}}.
\bibitem[{Meng et~al.(2022)Meng, Li, Xu, Li, and Xia}]{meng2022weakly}
\bibinfo{author}{C.~Meng}, \bibinfo{author}{Y.~Li}, \bibinfo{author}{Y.~Xu}, \bibinfo{author}{N.~Li}, \bibinfo{author}{K.~Xia},
\newblock \bibinfo{title}{A weakly supervised framework for 2{D}/3{D} vascular registration oriented to incomplete 2{D} blood vessels},
\newblock \bibinfo{journal}{IEEE Trans. Med. Robot. Bionics} \bibinfo{volume}{4} (\bibinfo{year}{2022}) \bibinfo{pages}{381--390}.
\bibitem[{Abdelaziz et~al.(2021)Abdelaziz, Tian, Hamady, Yang, and Temelkuran}]{abdelaziz2021x}
\bibinfo{author}{M.~E. Abdelaziz}, \bibinfo{author}{L.~Tian}, \bibinfo{author}{M.~Hamady}, \bibinfo{author}{G.-Z. Yang}, \bibinfo{author}{B.~Temelkuran},
\newblock \bibinfo{title}{X-ray to {MR}: {T}he progress of flexible instruments for endovascular navigation},
\newblock \bibinfo{journal}{Prog. Biomed. Eng.} \bibinfo{volume}{3} (\bibinfo{year}{2021}) \bibinfo{pages}{032004}.
\bibitem[{Eagleton(2022)}]{eagleton2022updates}
\bibinfo{author}{M.~J. Eagleton},
\newblock \bibinfo{title}{Updates in endovascular procedural navigation},
\newblock \bibinfo{journal}{Can. J. Cardiol.} \bibinfo{volume}{38} (\bibinfo{year}{2022}) \bibinfo{pages}{662--671}.
\bibitem[{Huang et~al.(2024)}]{huang2024real}
\bibinfo{author}{D.-X. Huang}, et~al.,
\newblock \bibinfo{title}{Real-time 2{D}/3{D} registration via {CNN} regression and centroid alignment},
\newblock \bibinfo{journal}{IEEE Trans. Autom. Sci. Eng.}  (\bibinfo{year}{2024}). \bibinfo{note}{{D}OI:{\color{blue} \href{http://dx.doi.org/10.1109/TASE.2023.3345927}{10.1109/TASE.2023.3345927}}}.
\bibitem[{Zhu et~al.(2023)}]{zhu20233d}
\bibinfo{author}{J.~Zhu}, et~al.,
\newblock \bibinfo{title}{3{D}/2{D} vessel registration based on {M}onte {C}arlo tree search and manifold regularization},
\newblock \bibinfo{journal}{IEEE Trans. Med. Imaging}  (\bibinfo{year}{2023}). \bibinfo{note}{{D}OI:{\color{blue} \href{http://dx.doi.org/10.1109/TMI.2023.3347896}{10.1109/TMI.2023.3347896}}}.
\bibitem[{Ma et~al.(2021)}]{ma2021self}
\bibinfo{author}{Y.~Ma}, et~al.,
\newblock \bibinfo{title}{Self-supervised vessel segmentation via adversarial learning},
\newblock in: \bibinfo{booktitle}{Proc. ICCV}, \bibinfo{year}{2021}, pp. \bibinfo{pages}{7536--7545}.
\bibitem[{Xia et~al.(2019)}]{xia2019vessel}
\bibinfo{author}{S.~Xia}, et~al.,
\newblock \bibinfo{title}{Vessel segmentation of {X}-ray coronary angiographic image sequence},
\newblock \bibinfo{journal}{IEEE Trans. Biomed. Eng.} \bibinfo{volume}{67} (\bibinfo{year}{2019}) \bibinfo{pages}{1338--1348}.
\bibitem[{Huang et~al.(2026)}]{huang2026vasomim}
\bibinfo{author}{D.-X. Huang}, et~al.,
\newblock \bibinfo{title}{{VasoMIM}: Vascular anatomy-aware masked image modeling for vessel segmentation},
\newblock in: \bibinfo{booktitle}{Proc. AAAI}, \bibinfo{year}{2026}.
\bibitem[{Frangi et~al.(1998)Frangi, Niessen, Vincken, and Viergever}]{frangi1998multiscale}
\bibinfo{author}{A.~F. Frangi}, \bibinfo{author}{W.~J. Niessen}, \bibinfo{author}{K.~L. Vincken}, \bibinfo{author}{M.~A. Viergever},
\newblock \bibinfo{title}{Multiscale vessel enhancement filtering},
\newblock in: \bibinfo{booktitle}{Proc. MICCAI}, \bibinfo{year}{1998}, pp. \bibinfo{pages}{130--137}.
\bibitem[{Manniesing et~al.(2006)Manniesing, Viergever, and Niessen}]{manniesing2006vessel}
\bibinfo{author}{R.~Manniesing}, \bibinfo{author}{M.~A. Viergever}, \bibinfo{author}{W.~J. Niessen},
\newblock \bibinfo{title}{Vessel enhancing diffusion: A scale space representation of vessel structures},
\newblock \bibinfo{journal}{Med. Image Anal.} \bibinfo{volume}{10} (\bibinfo{year}{2006}) \bibinfo{pages}{815--825}.
\bibitem[{Cervantes-Sanchez et~al.(2016)}]{cervantes2016segmentation}
\bibinfo{author}{Cervantes-Sanchez}, et~al.,
\newblock \bibinfo{title}{Segmentation of coronary angiograms using {Gabor} filters and {Boltzmann} univariate marginal distribution algorithm},
\newblock \bibinfo{journal}{Comput. Intell. Neurosci.} \bibinfo{volume}{2016} (\bibinfo{year}{2016}).
\bibitem[{Jiang et~al.(2013)}]{jiang2013region}
\bibinfo{author}{H.~Jiang}, et~al.,
\newblock \bibinfo{title}{A region growing vessel segmentation algorithm based on spectrum information},
\newblock \bibinfo{journal}{Comput. Math. Methods Med.} \bibinfo{volume}{2013} (\bibinfo{year}{2013}) \bibinfo{pages}{743870}.
\bibitem[{Zeng et~al.(2018)}]{zeng2018automatic}
\bibinfo{author}{Y.-Z. Zeng}, et~al.,
\newblock \bibinfo{title}{Automatic liver vessel segmentation using {3D} region growing and hybrid active contour model},
\newblock \bibinfo{journal}{Comput. Biol. Med.} \bibinfo{volume}{97} (\bibinfo{year}{2018}) \bibinfo{pages}{63--73}.
\bibitem[{Sangsefidi et~al.(2018)Sangsefidi, Foruzan, and Dolati}]{sangsefidi2018balancing}
\bibinfo{author}{N.~Sangsefidi}, \bibinfo{author}{A.~H. Foruzan}, \bibinfo{author}{A.~Dolati},
\newblock \bibinfo{title}{Balancing the data term of graph-cuts algorithm to improve segmentation of hepatic vascular structures},
\newblock \bibinfo{journal}{Comput. Biol. Med.} \bibinfo{volume}{93} (\bibinfo{year}{2018}) \bibinfo{pages}{117--126}.
\bibitem[{Wang et~al.(2020)}]{wang2020tensor}
\bibinfo{author}{C.~Wang}, et~al.,
\newblock \bibinfo{title}{Tensor-cut: A tensor-based graph-cut blood vessel segmentation method and its application to renal artery segmentation},
\newblock \bibinfo{journal}{Med. Image Anal.} \bibinfo{volume}{60} (\bibinfo{year}{2020}) \bibinfo{pages}{101623}.
\bibitem[{Li et~al.(2020)Li, Bian, Zhou, Xie, Ni, and Hou}]{li2020cau}
\bibinfo{author}{R.-Q. Li}, \bibinfo{author}{G.-B. Bian}, \bibinfo{author}{X.-H. Zhou}, \bibinfo{author}{X.~Xie}, \bibinfo{author}{Z.-L. Ni}, \bibinfo{author}{Z.~Hou},
\newblock \bibinfo{title}{{CAU-net}: A novel convolutional neural network for coronary artery segmentation in digital subtraction angiography},
\newblock in: \bibinfo{booktitle}{Proc. ICONIP}, \bibinfo{year}{2020}, pp. \bibinfo{pages}{185--196}.
\bibitem[{Minaee et~al.(2021)Minaee, Boykov, Porikli, Plaza, Kehtarnavaz, and Terzopoulos}]{minaee2021image}
\bibinfo{author}{S.~Minaee}, \bibinfo{author}{Y.~Boykov}, \bibinfo{author}{F.~Porikli}, \bibinfo{author}{A.~Plaza}, \bibinfo{author}{N.~Kehtarnavaz}, \bibinfo{author}{D.~Terzopoulos},
\newblock \bibinfo{title}{Image segmentation using deep learning: A survey},
\newblock \bibinfo{journal}{IEEE Trans. Pattern Anal. Mach. Intell.} \bibinfo{volume}{44} (\bibinfo{year}{2021}) \bibinfo{pages}{3523--3542}.
\bibitem[{Chen et~al.(2022)}]{chen2022recent}
\bibinfo{author}{X.~Chen}, et~al.,
\newblock \bibinfo{title}{Recent advances and clinical applications of deep learning in medical image analysis},
\newblock \bibinfo{journal}{Med. Image Anal.} \bibinfo{volume}{79} (\bibinfo{year}{2022}) \bibinfo{pages}{102444}.
\bibitem[{Xiao et~al.(2026)}]{xiao2026reason}
\bibinfo{author}{N.-F. Xiao}, et~al.,
\newblock \bibinfo{title}{{REASON}: Probability map-guided dual-branch fusion framework for gastric content assessment},
\newblock \bibinfo{journal}{Biomed. Signal Process. Control}  (\bibinfo{year}{2026}).
\bibitem[{Ronneberger et~al.(2015)Ronneberger, Fischer, and Brox}]{ronneberger2015u}
\bibinfo{author}{O.~Ronneberger}, \bibinfo{author}{P.~Fischer}, \bibinfo{author}{T.~Brox},
\newblock \bibinfo{title}{U-{N}et: {C}onvolutional networks for biomedical image segmentation},
\newblock in: \bibinfo{booktitle}{Proc. MICCAI}, \bibinfo{year}{2015}, pp. \bibinfo{pages}{234--241}.
\bibitem[{Zhou et~al.(2019)Zhou, Siddiquee, Tajbakhsh, and Liang}]{zhou2019unet++}
\bibinfo{author}{Z.~Zhou}, \bibinfo{author}{M.~M.~R. Siddiquee}, \bibinfo{author}{N.~Tajbakhsh}, \bibinfo{author}{J.~Liang},
\newblock \bibinfo{title}{U{N}et++: {R}edesigning skip connections to exploit multiscale features in image segmentation},
\newblock \bibinfo{journal}{IEEE Trans. Med. Imaging} \bibinfo{volume}{39} (\bibinfo{year}{2019}) \bibinfo{pages}{1856--1867}.
\bibitem[{Schlemper et~al.(2019)}]{schlemper2019attention}
\bibinfo{author}{J.~Schlemper}, et~al.,
\newblock \bibinfo{title}{Attention gated networks: {L}earning to leverage salient regions in medical images},
\newblock \bibinfo{journal}{Med. Image Anal.} \bibinfo{volume}{53} (\bibinfo{year}{2019}) \bibinfo{pages}{197--207}.
\bibitem[{Huang et~al.(2026)}]{huang2026mosformer}
\bibinfo{author}{D.-X. Huang}, et~al.,
\newblock \bibinfo{title}{{MOSformer}: Momentum encoder-based inter-slice fusion transformer for medical image segmentation},
\newblock \bibinfo{journal}{Biomim. Intell. Robot.}  (\bibinfo{year}{2026}) \bibinfo{pages}{100298}.
\bibitem[{Mou et~al.(2021)}]{mou2021cs2}
\bibinfo{author}{L.~Mou}, et~al.,
\newblock \bibinfo{title}{C{S}$^2$-{N}et: Deep learning segmentation of curvilinear structures in medical imaging},
\newblock \bibinfo{journal}{Med. Image Anal.} \bibinfo{volume}{67} (\bibinfo{year}{2021}) \bibinfo{pages}{101874}.
\bibitem[{Liu et~al.(2022)}]{liu2022full}
\bibinfo{author}{W.~Liu}, et~al.,
\newblock \bibinfo{title}{Full-resolution network and dual-threshold iteration for retinal vessel and coronary angiograph segmentation},
\newblock \bibinfo{journal}{IEEE J. Biomed. Health Inform.} \bibinfo{volume}{26} (\bibinfo{year}{2022}) \bibinfo{pages}{4623--4634}.
\bibitem[{Li et~al.(2023)}]{li2023global}
\bibinfo{author}{Y.~Li}, et~al.,
\newblock \bibinfo{title}{Global transformer and dual local attention network via deep-shallow hierarchical feature fusion for retinal vessel segmentation},
\newblock \bibinfo{journal}{IEEE Trans. Cybern.} \bibinfo{volume}{53} (\bibinfo{year}{2023}) \bibinfo{pages}{5826--5839}.
\bibitem[{Dosovitskiy et~al.(2021)}]{dosovitskiy2020image}
\bibinfo{author}{A.~Dosovitskiy}, et~al.,
\newblock \bibinfo{title}{An image is worth 16x16 words: Transformers for image recognition at scale},
\newblock in: \bibinfo{booktitle}{Proc. ICLR}, \bibinfo{year}{2021}.
\bibitem[{Liu et~al.(2021)}]{liu2021swin}
\bibinfo{author}{Z.~Liu}, et~al.,
\newblock \bibinfo{title}{Swin {T}ransformer: Hierarchical vision transformer using shifted windows},
\newblock in: \bibinfo{booktitle}{Proc. ICCV}, \bibinfo{year}{2021}, pp. \bibinfo{pages}{10012--10022}.
\bibitem[{Chen et~al.(2024)}]{chen2024transunet}
\bibinfo{author}{J.~Chen}, et~al.,
\newblock \bibinfo{title}{{TransUNet}: Rethinking the {U-Net} architecture design for medical image segmentation through the lens of transformers},
\newblock \bibinfo{journal}{Med. Image Anal.} \bibinfo{volume}{97} (\bibinfo{year}{2024}) \bibinfo{pages}{103280}.
\bibitem[{Guo et~al.(2022)}]{guo2022cmt}
\bibinfo{author}{J.~Guo}, et~al.,
\newblock \bibinfo{title}{{CMT}: Convolutional neural networks meet vision transformers},
\newblock in: \bibinfo{booktitle}{Proc. CVPR}, \bibinfo{year}{2022}, pp. \bibinfo{pages}{12175--12185}.
\bibitem[{Heidari et~al.(2023)}]{heidari2023hiformer}
\bibinfo{author}{M.~Heidari}, et~al.,
\newblock \bibinfo{title}{Hi{F}ormer: {H}ierarchical multi-scale representations using transformers for medical image segmentation},
\newblock in: \bibinfo{booktitle}{Proc. WACV}, \bibinfo{year}{2023}, pp. \bibinfo{pages}{6202--6212}.
\bibitem[{Kuang et~al.(2024)}]{kuang2024hybrid}
\bibinfo{author}{H.~Kuang}, et~al.,
\newblock \bibinfo{title}{Hybrid {CNN}-{T}ransformer network with circular feature interaction for acute ischemic stroke lesion segmentation on non-contrast {CT} scans},
\newblock \bibinfo{journal}{IEEE Trans. Med. Imaging}  (\bibinfo{year}{2024}). \bibinfo{note}{{D}OI:{\color{blue} \href{http://dx.doi.org/10.1109/TMI.2024.3362879}{10.1109/TMI.2024.3362879}}}.
\bibitem[{Heidbuchel et~al.(2014)}]{heidbuchel2014practical}
\bibinfo{author}{H.~Heidbuchel}, et~al.,
\newblock \bibinfo{title}{Practical ways to reduce radiation dose for patients and staff during device implantations and electrophysiological procedures},
\newblock \bibinfo{journal}{Europace} \bibinfo{volume}{16} (\bibinfo{year}{2014}) \bibinfo{pages}{946--964}.
\bibitem[{Adams and Bischof(1994)}]{adams1994seeded}
\bibinfo{author}{R.~Adams}, \bibinfo{author}{L.~Bischof},
\newblock \bibinfo{title}{Seeded region growing},
\newblock \bibinfo{journal}{IEEE Trans. Pattern Anal. Mach. Intell.} \bibinfo{volume}{16} (\bibinfo{year}{1994}) \bibinfo{pages}{641--647}.
\bibitem[{Rodrigues et~al.(2020)Rodrigues, Conci, and Liatsis}]{rodrigues2020element}
\bibinfo{author}{E.~O. Rodrigues}, \bibinfo{author}{A.~Conci}, \bibinfo{author}{P.~Liatsis},
\newblock \bibinfo{title}{{ELEMENT}: Multi-modal retinal vessel segmentation based on a coupled region growing and machine learning approach},
\newblock \bibinfo{journal}{IEEE J. Biomed. Health Inform.} \bibinfo{volume}{24} (\bibinfo{year}{2020}) \bibinfo{pages}{3507--3519}.
\bibitem[{Taghizadeh~Dehkordi et~al.(2014)Taghizadeh~Dehkordi, Doost~Hoseini, Sadri, and Soltanianzadeh}]{taghizadeh2014local}
\bibinfo{author}{M.~Taghizadeh~Dehkordi}, \bibinfo{author}{A.~M. Doost~Hoseini}, \bibinfo{author}{S.~Sadri}, \bibinfo{author}{H.~Soltanianzadeh},
\newblock \bibinfo{title}{Local feature fitting active contour for segmenting vessels in angiograms},
\newblock \bibinfo{journal}{IET Comput. Vis.} \bibinfo{volume}{8} (\bibinfo{year}{2014}) \bibinfo{pages}{161--170}.
\bibitem[{Memari et~al.(2019)Memari, Ramli, Saripan, Mashohor, and Moghbel}]{memari2019retinal}
\bibinfo{author}{N.~Memari}, \bibinfo{author}{A.~R. Ramli}, \bibinfo{author}{M.~I.~B. Saripan}, \bibinfo{author}{S.~Mashohor}, \bibinfo{author}{M.~Moghbel},
\newblock \bibinfo{title}{Retinal blood vessel segmentation by using matched filtering and fuzzy {C}-means clustering with integrated level set method for diabetic retinopathy assessment},
\newblock \bibinfo{journal}{J. Med. Biol. Eng.} \bibinfo{volume}{39} (\bibinfo{year}{2019}) \bibinfo{pages}{713--731}.
\bibitem[{Kim et~al.(2022)Kim, Oh, and Ye}]{kim2022diffusion}
\bibinfo{author}{B.~Kim}, \bibinfo{author}{Y.~Oh}, \bibinfo{author}{J.~C. Ye},
\newblock \bibinfo{title}{Diffusion adversarial representation learning for self-supervised vessel segmentation},
\newblock \bibinfo{journal}{arXiv:2209.14566}  (\bibinfo{year}{2022}).
\bibitem[{Gu et~al.(2019)}]{gu2019net}
\bibinfo{author}{Z.~Gu}, et~al.,
\newblock \bibinfo{title}{{CE-Net}: {C}ontext encoder network for 2{D} medical image segmentation},
\newblock \bibinfo{journal}{IEEE Trans. Med. Imaging} \bibinfo{volume}{38} (\bibinfo{year}{2019}) \bibinfo{pages}{2281--2292}.
\bibitem[{Xie et~al.(2021)Xie, Wang, Yu, Anandkumar, Alvarez, and Luo}]{xie2021segformer}
\bibinfo{author}{E.~Xie}, \bibinfo{author}{W.~Wang}, \bibinfo{author}{Z.~Yu}, \bibinfo{author}{A.~Anandkumar}, \bibinfo{author}{J.~M. Alvarez}, \bibinfo{author}{P.~Luo},
\newblock \bibinfo{title}{{SegFormer}: Simple and efficient design for semantic segmentation with transformers},
\newblock in: \bibinfo{booktitle}{Proc. NeurIPS}, volume~\bibinfo{volume}{34}, \bibinfo{year}{2021}, pp. \bibinfo{pages}{12077--12090}.
\bibitem[{Wang et~al.(2022)Wang, Cao, Wang, and Zaiane}]{wang2022uctransnet}
\bibinfo{author}{H.~Wang}, \bibinfo{author}{P.~Cao}, \bibinfo{author}{J.~Wang}, \bibinfo{author}{O.~R. Zaiane},
\newblock \bibinfo{title}{{UCTransNet}: Rethinking the skip connections in {U-Net} from a channel-wise perspective with transformer},
\newblock in: \bibinfo{booktitle}{Proc. AAAI}, volume~\bibinfo{volume}{36}, \bibinfo{year}{2022}, pp. \bibinfo{pages}{2441--2449}.
\bibitem[{Gonzales and Wintz(1987)}]{gonzales1987digital}
\bibinfo{author}{R.~C. Gonzales}, \bibinfo{author}{P.~Wintz}, \bibinfo{title}{Digital image processing}, \bibinfo{publisher}{Addison-Wesley Longman Publishing Co., Inc.}, \bibinfo{year}{1987}.
\bibitem[{Cui et~al.(2025)Cui, Zamir, Khan, Knoll, Shah, and Khan}]{cui2025adair}
\bibinfo{author}{Y.~Cui}, \bibinfo{author}{S.~W. Zamir}, \bibinfo{author}{S.~Khan}, \bibinfo{author}{A.~Knoll}, \bibinfo{author}{M.~Shah}, \bibinfo{author}{F.~Khan},
\newblock \bibinfo{title}{{AdaIR}: Adaptive all-in-one image restoration via frequency mining and modulation},
\newblock in: \bibinfo{booktitle}{Proc. ICLR}, \bibinfo{year}{2025}, pp. \bibinfo{pages}{101306--101327}.
\bibitem[{Azad et~al.(2021)Azad, Bozorgpour, Asadi-Aghbolaghi, Merhof, and Escalera}]{azad2021deep}
\bibinfo{author}{R.~Azad}, \bibinfo{author}{A.~Bozorgpour}, \bibinfo{author}{M.~Asadi-Aghbolaghi}, \bibinfo{author}{D.~Merhof}, \bibinfo{author}{S.~Escalera},
\newblock \bibinfo{title}{Deep frequency re-calibration {U-Net} for medical image segmentation},
\newblock in: \bibinfo{booktitle}{Proc. ICCVW}, \bibinfo{year}{2021}, pp. \bibinfo{pages}{3274--3283}.
\bibitem[{Chi et~al.(2020)Chi, Jiang, and Mu}]{chi2020fast}
\bibinfo{author}{L.~Chi}, \bibinfo{author}{B.~Jiang}, \bibinfo{author}{Y.~Mu},
\newblock \bibinfo{title}{Fast {Fourier} convolution},
\newblock in: \bibinfo{booktitle}{Proc. NeurIPS}, volume~\bibinfo{volume}{33}, \bibinfo{year}{2020}, pp. \bibinfo{pages}{4479--4488}.
\bibitem[{Guibas et~al.(2021)Guibas, Mardani, Li, Tao, Anandkumar, and Catanzaro}]{guibas2021adaptive}
\bibinfo{author}{J.~Guibas}, \bibinfo{author}{M.~Mardani}, \bibinfo{author}{Z.~Li}, \bibinfo{author}{A.~Tao}, \bibinfo{author}{A.~Anandkumar}, \bibinfo{author}{B.~Catanzaro},
\newblock \bibinfo{title}{Adaptive {Fourier} neural operators: Efficient token mixers for transformers},
\newblock \bibinfo{journal}{arXiv:2111.13587}  (\bibinfo{year}{2021}).
\bibitem[{Rao et~al.(2023)Rao, Zhao, Zhu, Zhou, and Lu}]{rao2023gfnet}
\bibinfo{author}{Y.~Rao}, \bibinfo{author}{W.~Zhao}, \bibinfo{author}{Z.~Zhu}, \bibinfo{author}{J.~Zhou}, \bibinfo{author}{J.~Lu},
\newblock \bibinfo{title}{{GFNet}: Global filter networks for visual recognition},
\newblock \bibinfo{journal}{IEEE Trans. Pattern Anal. Mach. Intell.} \bibinfo{volume}{45} (\bibinfo{year}{2023}) \bibinfo{pages}{10960--10973}.
\bibitem[{Huang et~al.(2023)Huang, Zhang, Lan, Zha, Lu, and Guo}]{huang2023adaptive}
\bibinfo{author}{Z.~Huang}, \bibinfo{author}{Z.~Zhang}, \bibinfo{author}{C.~Lan}, \bibinfo{author}{Z.-J. Zha}, \bibinfo{author}{Y.~Lu}, \bibinfo{author}{B.~Guo},
\newblock \bibinfo{title}{Adaptive frequency filters as efficient global token mixers},
\newblock in: \bibinfo{booktitle}{Proc. ICCV}, \bibinfo{year}{2023}, pp. \bibinfo{pages}{6049--6059}.
\bibitem[{Huang et~al.(2021)Huang, Zhou, Chen, Chen, and Lan}]{huang2021medical}
\bibinfo{author}{Y.~Huang}, \bibinfo{author}{C.~Zhou}, \bibinfo{author}{L.~Chen}, \bibinfo{author}{J.~Chen}, \bibinfo{author}{S.~Lan},
\newblock \bibinfo{title}{Medical frequency domain learning: Consider inter-class and intra-class frequency for medical image segmentation and classification},
\newblock in: \bibinfo{booktitle}{Proc. BIBM}, \bibinfo{year}{2021}, pp. \bibinfo{pages}{897--904}.
\bibitem[{Li et~al.(2023)Li, Zhou, He, Zhao, and Tian}]{li2023global2}
\bibinfo{author}{P.~Li}, \bibinfo{author}{R.~Zhou}, \bibinfo{author}{J.~He}, \bibinfo{author}{S.~Zhao}, \bibinfo{author}{Y.~Tian},
\newblock \bibinfo{title}{A global-frequency-domain network for medical image segmentation},
\newblock \bibinfo{journal}{Comput. Biol. Med.} \bibinfo{volume}{164} (\bibinfo{year}{2023}) \bibinfo{pages}{107290}.
\bibitem[{Fu et~al.(2020)Fu, Liu, Jiang, Li, Bao, and Lu}]{fu2020scene}
\bibinfo{author}{J.~Fu}, \bibinfo{author}{J.~Liu}, \bibinfo{author}{J.~Jiang}, \bibinfo{author}{Y.~Li}, \bibinfo{author}{Y.~Bao}, \bibinfo{author}{H.~Lu},
\newblock \bibinfo{title}{Scene segmentation with dual relation-aware attention network},
\newblock \bibinfo{journal}{IEEE Trans. Neural Netw. Learn. Syst.} \bibinfo{volume}{32} (\bibinfo{year}{2020}) \bibinfo{pages}{2547--2560}.
\bibitem[{Li et~al.(2022)Li, Zhang, Cui, Lei, Kuang, and Zhang}]{li2022dual}
\bibinfo{author}{Y.~Li}, \bibinfo{author}{Y.~Zhang}, \bibinfo{author}{W.~Cui}, \bibinfo{author}{B.~Lei}, \bibinfo{author}{X.~Kuang}, \bibinfo{author}{T.~Zhang},
\newblock \bibinfo{title}{Dual encoder-based dynamic-channel graph convolutional network with edge enhancement for retinal vessel segmentation},
\newblock \bibinfo{journal}{IEEE Trans. Med. Imaging} \bibinfo{volume}{41} (\bibinfo{year}{2022}) \bibinfo{pages}{1975--1989}.
\bibitem[{Hu et~al.(2019)Hu, Shen, Albanie, Sun, and Wu}]{hu2019squeeze}
\bibinfo{author}{J.~Hu}, \bibinfo{author}{L.~Shen}, \bibinfo{author}{S.~Albanie}, \bibinfo{author}{G.~Sun}, \bibinfo{author}{E.~Wu},
\newblock \bibinfo{title}{Squeeze-and-{E}xcitation networks},
\newblock \bibinfo{journal}{IEEE Trans. Pattern Anal. Mach. Intell.} \bibinfo{volume}{42} (\bibinfo{year}{2019}) \bibinfo{pages}{2011--2023}.
\bibitem[{Woo et~al.(2018)Woo, Park, Lee, and Kweon}]{woo2018cbam}
\bibinfo{author}{S.~Woo}, \bibinfo{author}{J.~Park}, \bibinfo{author}{J.-Y. Lee}, \bibinfo{author}{I.~S. Kweon},
\newblock \bibinfo{title}{{CBAM}: Convolutional block attention module},
\newblock in: \bibinfo{booktitle}{Proc. ECCV}, \bibinfo{year}{2018}, pp. \bibinfo{pages}{3--19}.
\bibitem[{Wang et~al.(2020)Wang, Wu, Zhu, Li, Zuo, and Hu}]{wang2020eca}
\bibinfo{author}{Q.~Wang}, \bibinfo{author}{B.~Wu}, \bibinfo{author}{P.~Zhu}, \bibinfo{author}{P.~Li}, \bibinfo{author}{W.~Zuo}, \bibinfo{author}{Q.~Hu},
\newblock \bibinfo{title}{{ECA-Net}: Efficient channel attention for deep convolutional neural networks},
\newblock in: \bibinfo{booktitle}{Proc. CVPR}, \bibinfo{year}{2020}, pp. \bibinfo{pages}{11534--11542}.
\bibitem[{Guo et~al.(2021)Guo, Szemenyei, Hu, Wang, Zhou, and Yi}]{guo2021channel}
\bibinfo{author}{C.~Guo}, \bibinfo{author}{M.~Szemenyei}, \bibinfo{author}{Y.~Hu}, \bibinfo{author}{W.~Wang}, \bibinfo{author}{W.~Zhou}, \bibinfo{author}{Y.~Yi},
\newblock \bibinfo{title}{Channel attention residual {U-Net} for retinal vessel segmentation},
\newblock in: \bibinfo{booktitle}{Proc. ICASSP}, \bibinfo{year}{2021}, pp. \bibinfo{pages}{1185--1189}.
\bibitem[{Cooley and Tukey(1965)}]{cooley1965algorithm}
\bibinfo{author}{J.~W. Cooley}, \bibinfo{author}{J.~W. Tukey},
\newblock \bibinfo{title}{An algorithm for the machine calculation of complex {Fourier} series},
\newblock \bibinfo{journal}{Math. Comput.} \bibinfo{volume}{19} (\bibinfo{year}{1965}) \bibinfo{pages}{297--301}.
\bibitem[{He et~al.(2016)He, Zhang, Ren, and Sun}]{he2016deep}
\bibinfo{author}{K.~He}, \bibinfo{author}{X.~Zhang}, \bibinfo{author}{S.~Ren}, \bibinfo{author}{J.~Sun},
\newblock \bibinfo{title}{Deep residual learning for image recognition},
\newblock in: \bibinfo{booktitle}{Proc. CVPR}, \bibinfo{year}{2016}, pp. \bibinfo{pages}{770--778}.
\bibitem[{Wang et~al.(2023)Wang, Jiang, Zhong, and Liu}]{wang2023spatial}
\bibinfo{author}{C.~Wang}, \bibinfo{author}{J.~Jiang}, \bibinfo{author}{Z.~Zhong}, \bibinfo{author}{X.~Liu},
\newblock \bibinfo{title}{Spatial-frequency mutual learning for face super-resolution},
\newblock in: \bibinfo{booktitle}{Proc. CVPR}, \bibinfo{year}{2023}, pp. \bibinfo{pages}{22356--22366}.
\bibitem[{Paszke et~al.(2019)}]{paszke2019pytorch}
\bibinfo{author}{A.~Paszke}, et~al.,
\newblock \bibinfo{title}{{PyTorch}: An imperative style, high-performance deep learning library},
\newblock in: \bibinfo{booktitle}{Proc. NeurIPS}, volume~\bibinfo{volume}{32}, \bibinfo{year}{2019}.
\bibitem[{Zhu et~al.(2019)Zhu, Xu, Bai, Huang, and Bai}]{zhu2019asymmetric}
\bibinfo{author}{Z.~Zhu}, \bibinfo{author}{M.~Xu}, \bibinfo{author}{S.~Bai}, \bibinfo{author}{T.~Huang}, \bibinfo{author}{X.~Bai},
\newblock \bibinfo{title}{Asymmetric non-local neural networks for semantic segmentation},
\newblock in: \bibinfo{booktitle}{Proc. ICCV}, \bibinfo{year}{2019}, pp. \bibinfo{pages}{593--602}.
\bibitem[{Kipf and Welling(2016)}]{kipf2016semi}
\bibinfo{author}{T.~N. Kipf}, \bibinfo{author}{M.~Welling},
\newblock \bibinfo{title}{Semi-supervised classification with graph convolutional networks},
\newblock \bibinfo{journal}{arXiv:1609.02907}  (\bibinfo{year}{2016}).
\bibitem[{Li et~al.(2020)Li, Yang, Zhao, Shen, Lin, and Liu}]{li2020spatial}
\bibinfo{author}{X.~Li}, \bibinfo{author}{Y.~Yang}, \bibinfo{author}{Q.~Zhao}, \bibinfo{author}{T.~Shen}, \bibinfo{author}{Z.~Lin}, \bibinfo{author}{H.~Liu},
\newblock \bibinfo{title}{Spatial pyramid based graph reasoning for semantic segmentation},
\newblock in: \bibinfo{booktitle}{Proc. CVPR}, \bibinfo{year}{2020}, pp. \bibinfo{pages}{8950--8959}.
\bibitem[{Yushkevich et~al.(2006)}]{yushkevich2006user}
\bibinfo{author}{P.~A. Yushkevich}, et~al.,
\newblock \bibinfo{title}{User-guided 3{D} active contour segmentation of anatomical structures: Significantly improved efficiency and reliability},
\newblock \bibinfo{journal}{NeuroImage} \bibinfo{volume}{31} (\bibinfo{year}{2006}) \bibinfo{pages}{1116--1128}.
\bibitem[{Cervantes-Sanchez et~al.(2019)Cervantes-Sanchez, Cruz-Aceves, Hernandez-Aguirre, Hernandez-Gonzalez, and Solorio-Meza}]{cervantes2019automatic}
\bibinfo{author}{F.~Cervantes-Sanchez}, \bibinfo{author}{I.~Cruz-Aceves}, \bibinfo{author}{A.~Hernandez-Aguirre}, \bibinfo{author}{M.~A. Hernandez-Gonzalez}, \bibinfo{author}{S.~E. Solorio-Meza},
\newblock \bibinfo{title}{Automatic segmentation of coronary arteries in {X}-ray angiograms using multiscale analysis and artificial neural networks},
\newblock \bibinfo{journal}{Appl. Sci.} \bibinfo{volume}{9} (\bibinfo{year}{2019}) \bibinfo{pages}{5507}.
\bibitem[{Popov et~al.(2024)}]{popov2024dataset}
\bibinfo{author}{M.~Popov}, et~al.,
\newblock \bibinfo{title}{Dataset for automatic region-based coronary artery disease diagnostics using {X}-ray angiography images},
\newblock \bibinfo{journal}{Sci. Data} \bibinfo{volume}{11} (\bibinfo{year}{2024}) \bibinfo{pages}{20}.
\bibitem[{Wang et~al.(2018)Wang, Girshick, Gupta, and He}]{wang2018non}
\bibinfo{author}{X.~Wang}, \bibinfo{author}{R.~Girshick}, \bibinfo{author}{A.~Gupta}, \bibinfo{author}{K.~He},
\newblock \bibinfo{title}{Non-local neural networks},
\newblock in: \bibinfo{booktitle}{Proc. CVPR}, \bibinfo{year}{2018}, pp. \bibinfo{pages}{7794--7803}.
\bibitem[{Li et~al.(2019)Li, Wang, Hu, and Yang}]{li2019selective}
\bibinfo{author}{X.~Li}, \bibinfo{author}{W.~Wang}, \bibinfo{author}{X.~Hu}, \bibinfo{author}{J.~Yang},
\newblock \bibinfo{title}{Selective kernel networks},
\newblock in: \bibinfo{booktitle}{Proc. CVPR}, \bibinfo{year}{2019}, pp. \bibinfo{pages}{510--519}.
\bibitem[{Qin et~al.(2021)Qin, Zhang, Wu, and Li}]{qin2021fcanet}
\bibinfo{author}{Z.~Qin}, \bibinfo{author}{P.~Zhang}, \bibinfo{author}{F.~Wu}, \bibinfo{author}{X.~Li},
\newblock \bibinfo{title}{{FcaNet}: Frequency channel attention networks},
\newblock in: \bibinfo{booktitle}{Proc. ICCV}, \bibinfo{year}{2021}, pp. \bibinfo{pages}{783--792}.
\bibitem[{Selvaraju et~al.(2017)Selvaraju, Cogswell, Das, Vedantam, Parikh, and Batra}]{selvaraju2017grad}
\bibinfo{author}{R.~R. Selvaraju}, \bibinfo{author}{M.~Cogswell}, \bibinfo{author}{A.~Das}, \bibinfo{author}{R.~Vedantam}, \bibinfo{author}{D.~Parikh}, \bibinfo{author}{D.~Batra},
\newblock \bibinfo{title}{Grad-{CAM}: Visual explanations from deep networks via gradient-based localization},
\newblock in: \bibinfo{booktitle}{Proc. ICCV}, \bibinfo{year}{2017}, pp. \bibinfo{pages}{618--626}.
\bibitem[{Shit et~al.(2021)}]{shit2021cldice}
\bibinfo{author}{S.~Shit}, et~al.,
\newblock \bibinfo{title}{{clDice}-a novel topology-preserving loss function for tubular structure segmentation},
\newblock in: \bibinfo{booktitle}{Proc. CVPR}, \bibinfo{year}{2021}, pp. \bibinfo{pages}{16560--16569}.

\end{thebibliography}
